\documentclass[twocolumn, superscriptaddress]{aastex631}
\usepackage{CJK}
\usepackage{url}
\usepackage{amsmath}
\usepackage[super]{nth}
\usepackage{microtype}
\usepackage{graphicx}
\usepackage{rotating}

\shorttitle{A Unified Spectroscopic and Photometric Model to Infer Surface Inhomogeneity}
\shortauthors{Plummer et al.}

\graphicspath{{./}{figures/}}

\begin{document}
\begin{CJK*}{UTF8}{gbsn}

\title{A Unified Spectroscopic and Photometric Model to Infer Surface Inhomogeneity: Application to Luhman 16B}

\author[0000-0002-4831-0329]{Michael K. Plummer}
\affiliation{The Ohio State University \\
Department of Astronomy \\
140 W. 18th Ave. \\
Columbus, OH 43210, USA}

\correspondingauthor{Michael K. Plummer}
\email{plummer.323@osu.edu}

\author[0000-0002-4361-8885]{Ji Wang (王吉)}
\affiliation{The Ohio State University \\
Department of Astronomy \\
140 W. 18th Ave. \\
Columbus, OH 43210, USA}

\begin{abstract}

Extremely large telescopes (ELTs) provide an opportunity to observe surface inhomogeneities for ultracool objects including M dwarfs, brown dwarfs (BDs), and gas giant planets via Doppler imaging and spectro-photometry techniques. These inhomogeneities can be caused by star spots, clouds, and vortices. Star spots and associated stellar flares play a significant role in habitability, either stifling life or catalyzing abiogenesis depending on the emission frequency, magnitude, and orientation. Clouds and vortices may be the source of spectral and photometric variability observed at the L/T transition of BDs and are expected in gas giant exoplanets. We develop a versatile analytical framework to model and infer surface inhomogeneities which can be applied to both spectroscopic and photometric data. This model is validated against a slew of numerical simulations. Using archival spectroscopic and photometric data, we infer star spot parameters (location, size, and contrast) and generate global surface maps for Luhman 16B (an early T dwarf and one of our solar system's nearest neighbors at a distance of $\approx 2 \ \rm{pc}$). We confirm previous findings that Luhman 16B's atmosphere is inhomogeneous with time-varying features. In addition, we provide tentative evidence of longer timescale atmospheric structures such as dark equatorial and bright mid-latitude to polar spots. These findings are discussed in the context of atmospheric circulation and dynamics for ultracool dwarfs. Our analytical model will be valuable in assessing the feasibility of using ELTs to study surface inhomogeneities of gas giant exoplanets and other ultracool objects.

\end{abstract}


\section{Introduction}\label{sec:Intro}

Imaging ultracool object surface features such as star spots, clouds, and vortices will help us to better understand habitability, climate, and atmospheric dynamics. The advent of 30-meter-class extremely large telescopes (ELTs) and the James Webb Space Telescope (JWST) present vast opportunities in the fields of ultracool object imaging. ELTs will be unique in providing high signal-to-noise (SNR) and high resolution spectral data while JWST will provide high SNR photometric and low-to-medium resolution $(R \leq 5000)$ spectral data. These developments will dramatically improve our ability to map the global surfaces of ultracool objects such as M dwarfs, brown dwarfs (BDs), and gas giant exoplanets.

\subsection{Star Spots and M Dwarfs}\label{ssec:MDwarfs}

\par M dwarf stars comprise the most numerous spectral class in the Milky Way. Due to M dwarfs' small size, they are attractive targets for studies attempting to find habitable terrestrial planets because orbiting planets create correspondingly larger and more easily detected signatures with both the transit and radial velocity (RV) methods. Small rocky planets may be over three times more common around M dwarfs than FGK stars \citep{Mulders2015b}. The overabundance was later confirmed in comparing RV and transiting surveys \citep{Tuomi2019,Sabotta2021,Zink2020}.

\par Despite the prevalence of planets around M stars, obstacles exist for declaring M dwarf stellar systems ideal for planetary habitability. M dwarfs' low temperatures (compared to G dwarf stars like our Sun), mean planets located at distances where liquid water could exist on a planet's surface, approximately $10 \%$ the distance between the Earth and the Sun, would experience extensive stellar activity including flares \citep{kasting93, kopparapu13}. Fast-rotating M dwarfs, and particularly cooler, later and fully-convective stars, are far more likely to flare than their hotter counterparts which possess radiative zones \citep{gunther20}. Furthermore, flares and even superflares have been observed in ultracool objects with spectral classes as cool as L5, demonstrating that intense magnetic activity continues beyond the M-class range \citep{Paudel2018,Paudel2020}. 

\par Flares and coronal mass ejections (CMEs) associated with star spots could potentially bathe nearby, likely tidally-locked, planets in X-Ray and ultraviolet (UV) radiation and charged particles, stripping the planets' nascent atmospheres. However, it remains uncertain whether this radiation will inhibit or catalyze the emergence and growth of burgeoning life \citep{scalo07,tarter07,zendejas10}. Laboratory experiments suggest that stellar flares could provide the UV light necessary to photochemically produce the prerequisite molecules for abiogenesis \citep{rimmer18}.

\par Understanding star spot coverage as a proxy for magnetic activity and stellar flares is necessary to reveal how these energetic events may influence the long-term habitability of planetary systems. Late M dwarfs have been observed to have star spots preferentially located within their polar latitudes which differs from the predominantly equatorial-forming sun spots \citep{strassmeier&rice98,jarvinen08,barnes15,barnes17}. TESS observations have also provided preliminary evidence of white-light flares preferentially occurring at polar latitudes (between $55^{\circ}$ and $81^{\circ}$) in fully convective M dwarf stars \citep{ilin21}. Future M dwarf imaging will be critical to confirming theories on star spot formation and migration, and will also be necessary to understand and quantify habitability in these star systems.

\subsection{Brown Dwarfs at the L/T Transition}\label{ssec:BrownDwarfs}

\par Lower in mass than M dwarfs, BDs are substellar objects below the hydrogen-burning minimum mass $(M \lesssim 70 \ M_J)$ \citep{Chabrier2000, Dupuy2017}. They gradually cool over billions of years from early post-formation temperatures consistent with late M dwarfs, through L, T, and Y substellar classes. Across this span, the spectral features transition quite dramatically in $J-K$ color from red, cloudy L Dwarfs to relatively blue, cloud-free mid-T Dwarfs due to potentially two primary mechanisms related to the decreasing temperature: (1) Clouds composed of refractory materials such as iron (Fe), corundum (Al$_2$O$_3$), enstatite (MgSiO$_3$), and forsterite (Mg$_2$SiO$_4$) condense and precipitate out over the L/T transition leading to these materials seemingly being sequestered below the photosphere (2) Carbon monoxide (CO) is reduced to methane (CH$_4$) resulting $J$ band brightening and $H$ and $K$ suppression due to the ensuing CH$_4$ absorption \citep{Tsuji1996,Jones&Tsuji1997,Noll2000,Allard2001,Marley2002,Burrows2006,Cushing2008}. 

BDs at the L/T transition appear to demonstrate cloud patchiness which may explain the source of variability. It was initially theorized that the condensation of the metal and silicate clouds along with robust tropospheric convection leads to gaps in cloud coverage \citep{ackerman&marley01,Burgasser2002,Marley2010}. More recently, it has been argued that a more likely scenario is that condensation results in patchy regions of varying cloud height and optical thickness \citep{Radigan2012,Apai2013,Buenzli2014}. Inhomogeneous atmospheres paired with the fact most BDs spin rapidly with average rotational speeds exceeding $20 \ \rm{km} \cdot \rm{s}^{-1}$ for mid-L Dwarfs, naturally leads to the conclusion that ultracool objects at the L/T transition will experience large photometric and spectroscopic variability \citep{Reimers2008}.

\par Previous observations indicated that, although objects at the L/T transition were variable, they were perhaps no more variable than other BDs \citep{Enoch2003,Buenzli2014,Metchev2015}. However, \citet{Radigan2014a} deduced that $39_{-14}^{+16} \%$ of L/T transition stars were highly variable ($>2 \%$) while also demonstrating strong variability to be rare for BDs outside this range. \citet{Eriksson2019} determined a similar value of $40_{-19}^{+32} \%$ for the proportion of strongly variable L/T transition BDs. 

\par Future opportunities exist to shed more light on the exact nature of the L/T transition and BD atmospheric dynamics. ELTs and the JWST will provide high quality data that could potentially be used to map the surfaces of numerous nearby BDs. These maps very well may reveal the existence of atmospheric structure including zonal jets and vortices as well as the true source of L/T transition variability. With this information, we can further constrain atmospheric formation, structure, and dynamics.

\par Not only will we be able to map these ultracool objects' surface temperatures, we may also be able to map chemical tracers via Doppler imaging. This could allow us detect regions of chemical disequilibrium (\S \ref{ssec:discusion/disequilibrium}) originating from vertical atmospheric flow. Studying such behavior across spectral classes can allow us to fully understand the development of thermal and hydrodynamic structures with ultracool objects.

\subsection{Clouds in Gas Giant Exoplanet Atmospheres}\label{ssec:introgasgiants}

\par Directly-imaged gas giant exoplanets typically form at large distances from their host stars and are weakly-illuminated, leading to targets similar to BDs in terms of observational opportunities. Like their more massive BD cousins, gas giant exoplanets possess clouds \citep{Madhusudhan2011,Currie2011,Currie2014,Marley2012,Skemer2014} and chemical disequilibrium \citep{Barman2011,Barman2015,Konopacky2013,Skemer2014,Currie2014,Lavie2017,Molliere2020,Wang2020}. 

\par High SNR and high spectral resolution data from ELTs will be essential to creating gas giant surface maps via Doppler imaging in the future. These maps may have the ability to determine the extent of cloud patchiness as well as the occurrence of planetary-scale structures such as zonal jet bands, vortices, and persistent storms. As is later discussed in \S \ref{ssec:discusion/disequilibrium}, Doppler imaging may also be used to identify regions of convective activity through chemical disequilibrium tracers. 

\subsection{Doppler Imaging}\label{ssec:DopplerImaging}

\par Doppler imagining is an important spectroscopic method for studying surface inhomogeneity. The history of Doppler imaging goes back almost seventy years when \citet{Deutsch1958} first used spectral line profile (LP) deviations to map the magnetic fields and abundance anomalies of Ap stars; this method was further formalized in \citet{Deutsch1970} and \citet{Falk1974}. Ap stars continued to be targeted for surveys mapping their spots and chemical compositions due the class possessing strong surface inhomogeneities with century-long lifetimes \citep{Khokhlova1976,Goncharskii1977,Goncharskii1982}. Following this work, \citet{Vogt&Penrod1983} conducted a ground-breaking study involving the Doppler imaging of the spotted star, HR1099. 

\par These early Doppler imaging efforts suffered from the fact that stellar surface imaging is an ill-posed problem, the results were often degenerate with more than one solution for a given set of data \citep{Piskunov1990}. Solutions to this problem arose in applying maximum entropy methods \citep{vogt87} and Tikhonov regularization \citep{Piskunov1990} to infer the smoothest stellar surfaces. 

\par Since these developments, studies have generated Doppler images of giants and sub-giants \citep{Donati1999,Strassmeier1998,Strassmeier1999a,Strassmeier1999b}, pre-main sequence stars \citep{Hatzes1995,strassmeier&rice98,Finociety2021}, K Dwarfs \citep{CollierCameron&Unruh1994,Barnes&CollierCameron2001a,Barnes2004,Zaire2021}, and even M dwarfs \citep{Barnes&CollierCameron2001b,Barnes2004,barnes17}. Notably, \citet{crossfield14} conducted the first Doppler imaging of a BD on the nearby binary system Luhman 16; these results were reprocessed and replicated in \citet{luger21a}. 

\subsection{Photometry and Low-Resolution Techniques}

\par Photometric and low-resolution spectroscopic (spectro-photometric) campaigns have led to improved understanding of BD and directly-imaged gas giant exoplanet atmospheres and clouds. Atmospheric structure can be investigated by monitoring ultracool object rotational modulation \citep{Buenzli2012,Buenzli15b,Buenzli15a,Apai2013,karalidi16,Yang2016,Schlawin2017,Biller2018,Zhou2020b,Manjavacas2021}. The Hubble Space Telescope (HST) Cloud Atlas program has focused on using these methods and performed extensive spectro-photometric studies of BDs, gas giants, and planetary-mass objects \citep{Lew2016,Lew2020a,Lew2020b,Manjavacas2018,Manjavacas2019a,Manjavacas2019b,Zhou2018,Zhou2019,Zhou2020,MilesPaez2019}. Moving forward, these methods can potentially be integrated with techniques such as Doppler imaging to create more complete surface maps for ultracool objects. The unified model presented in this paper seeks to address this need by combining photometric techniques with Doppler imaging data sets.

\subsection{Outline}

\par In this paper, we develop analytical and numerical frameworks for Doppler imaging ultracool objects using the Euler-Rodrigues formula (\S \ref{sec:Methods}). We internally validate our analytical method versus a numerical stellar/substellar surface model (\S \ref{sec:validationsection}). Our model is then externally validated against spectroscopic and photometric archival data from \citet{crossfield14} and \citet{Buenzli15a} (\S \ref{sec:luhmansection}). Based on our results, we discuss the implications for Doppler imaging of ultracool objects and future avenues of research (\S \ref{sec:discussion}).

\section{Methods}\label{sec:Methods}

\subsection{Least Squares Deconvolution}\label{ssec:LSDsection}

\par LPs have higher SNR than individual spectral lines and are therefore more effective in identifying surface inhomogeneities. One common method of computing LPs for spectra uses the Cross-Correlation Function (CCF); however, for fast-rotating objects, Least Squares Deconvolution (LSD) can be a mathematically rigorous technique. LSD is an attractive option for both spectroscopic and spectropolarimetric applications because it can create high-precision, averaged LPs, ideal for Bayesian inference and Doppler imaging \citep{kochokhov10}. 

\par To understand how LSD works, we must consider how the target's observed spectral lines are broadened through multiple processes including rotation and instrumentation. We can account for these processes mathematically by convolving the source spectrum $S$ with a broadening and instrument kernel, represented by $Z$. This results in the observed spectrum $Y^{0}$ determined as such,

\begin{equation}\label{eqn:ObservedSpectrum}
Y^{0} = S \ast Z
\end{equation}

where $\ast$ is the convolution operator. In Equation (\ref{eqn:ObservedSpectrum}), the convolution operator broadens the source spectrum, $S$, by the width of the broadening kernel, $Z$, creating a mathematical representation of the observed spectrum.

\par In most cases, we do not have perfect information and cannot directly observe the broadening kernel. The source spectrum is only estimated based on prior observations and models. The parameters we wish to determine are often innate to the source spectrum, but our primary observable is the observed spectrum, $Y^{0}$. We estimate the source spectrum with a template spectrum, $F$, which can be computed from various models based on known or estimated parameters such as effective temperature (T$_{eff}$), $\log$(g), and metallicity.  

\par We conducted sensitivity tests for both effective temperature and log(g) with a fixed template spectra (T$_{eff}$ = 1500 K, $\log$(g) $= 5.0$, $[Fe/H] = 0$), in the range used in \citet{crossfield14}. We created synthetic observed spectra with varying effective temperature ($\pm$ 50 K) and log(g) ($\pm$ 0.5); available spectra in this temperature regime did not have multiple metallicities available. This synthetic spectra was implemented, along with the fixed template, into our numerical model (see \S \ref{ssec:numericalmodel}) to create LPs with a spot of known latitude, longitude, contrast, and radius. We then retrieved these parameters via Bayesian inference. A temperature difference of 50 K resulted in the following  deviations in the retrieved parameters: latitude (1.24$\sigma$), longitude (0.57$\sigma$), radius (0.57$\sigma$), and contrast (0.19$\sigma$). A $\log$(g) difference of 0.5 resulted in the following:  latitude (1.68$\sigma$), longitude (0.096$\sigma$), radius (0.9$\sigma$), and contrast (0.16$\sigma$). As can be seen, longitude, radius, and especially contrast were fairly robust to temperature and gravity deviations while latitude was less robust. 

\par We can replace $S$ with $F$ in Equation (\ref{eqn:ObservedSpectrum}) as follows:

\begin{equation}\label{eqn:ObservedSpectrumF}
Y^{0} = F \ast Z = MZ.
\end{equation}

In this equation, $Y^{0}$ is comprised of $m$ elements with spacing dependent on the spectral sampling of the observational instrument. In practice, solving Equation \ref{eqn:ObservedSpectrumF} requires expanding $F$ into a line mask, $M$, via an $m \times n$ Toeplitz matrix as described in \citet{Donati1997}. This provides us with the following form:

\begin{equation}\label{eqn:InvertedSpectrumwithMask}
Z = M^{-1} \: Y^{0}.
\end{equation}

\par Similar to \citet{crossfield14}, we use LSD to convert the observed spectrum into an LP in RV space. These LPs can then be compared to either a mean LP from the observation or to a modeled LP to infer surface inhomogeneities. The specific LSD method used in this paper was adopted from \citet{Wang17}, \citet{Wang18}, and \citet{PaiAsnodkar22} and is best represented by the following:

\begin{equation}\label{eqn:LSDeqn}
Z = \left (M^{T} \: S^{2} \: M + \Lambda R \right )^{-1} \: M^{T} \: S^{2} \: Y^{0}
\end{equation}

where $Z$ is the computed LP comprised of $n$ data points and $S$ is an $m \times m$ diagonal error matrix with $S_{ii}$ elements equal to $1/\sigma_{ii}$ where $\sigma_{ii}$ is the measurement error at each epoch. $\Lambda$ is a scalar regularization parameter while $R$ is the associated regularization Toeplitz matrix which is outlined in \citet{donatelli&reichel14}.

\subsection{Analytical Model}\label{ssec:AnalyticalSubsection}

\subsubsection{Scheme}

\par Doppler tomography examines the changes in a target's LP during a transit of its viewable hemisphere. Our analytical method for modeling stellar and substellar surfaces derives inspiration from Doppler tomographic studies of eclipsing binary systems \citep{Albrecht2007,Wang18} and transiting exoplanets \citep{Hirano2011,Johnson2014}. Spectral deviations created by a transiting companion or surface inhomogeneity blocking a portion of the target's visible surface can be measured using a high-SNR technique such as LSD. 

\par
\begin{figure*}
\centering
\includegraphics[width=1\textwidth]{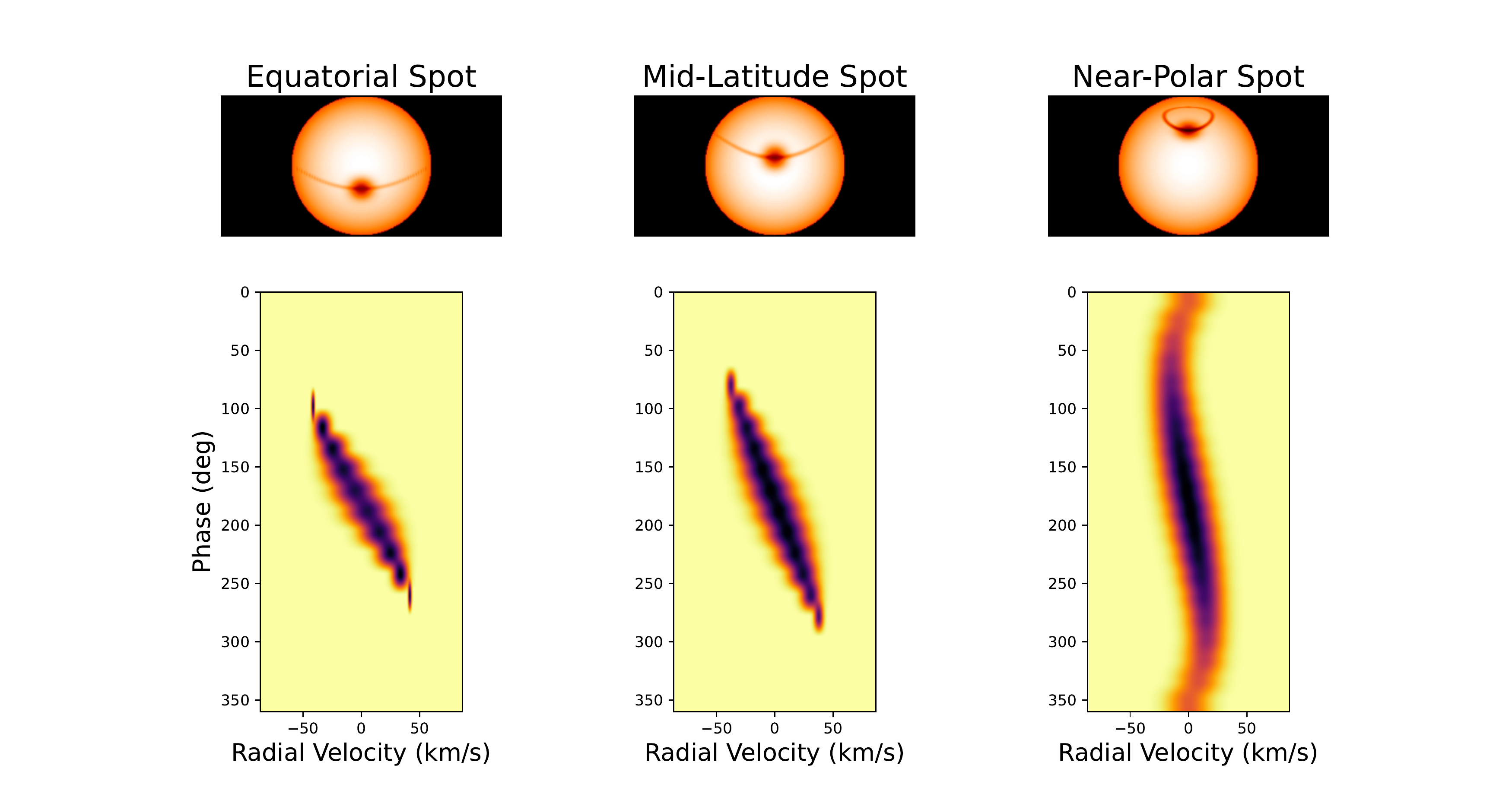}
\caption{\label{fig:DevPlotExample} Surface inhomogeneity trajectories and corresponding LP deviation plots for example stellar/substellar object with $\upsilon = 50 \ \ \rm{km}\cdot \rm{s}^{-1}$ and $i = 60^{\circ}$. (Left) Equatorial Spot ($l = 0^{\circ}$). (Middle) Mid-Latitude Spot ($l = 40^{\circ}$). (Right) Near-Polar Spot ($l = 75^{\circ}$). It should be noted that a transiting planet would create a linear signature versus the curved signatures seen in the figure.
}
\end{figure*}

 These spectral deviations create \textit{Doppler shadows} when they are plotted in RV space. With this technique, companion stars and planets will create linear shadows when they transit the target. In this article, we make the unique contribution of considering that surface inhomogeneities create curved Doppler shadows, if the target is inclined with respect to the viewer, as can be seen in Figure \ref{fig:DevPlotExample}. For the hemisphere inclined towards the viewer, higher latitudes spots are more extensive in their viewable phase.

\par We represent the unperturbed stellar/substellar disk as a rotationally-broadened LP. Surface inhomogeneities are modeled as Gaussian distributions whose width depends on radius and viewing angle. These Gaussian distributions are either subtracted from (dark spot) or added to (bright spot) the rotationally-broadened LP to create an analytical model for each epoch. An example of this process can be seen in Figure \ref{fig:analyticalmultispotexample} with dark spots subtracting from the rotationally-broadened LP.

\subsubsection{Unperturbed Line Profile}

\par We approximate the unperturbed LP with a rotational broadening kernel. Rotation is the dominant spectral line broadening mechanism for fast rotating astrophysical objects ($\upsilon \: sin \: i> 20 \ \rm{km}\cdot \rm{s}^{-1}$, where $\upsilon$ is the equatorial rotation speed of the object and $i$ is its inclination). 

\par A suitable expression for this rotational broadening kernel is described in \citet{gray22} and adapted by \citet{PaiAsnodkar22} as follows,

\begin{multline}\label{eqn:RotationalBroadEqn}
G(\upsilon_{i},t) = \\
c_{1}\left [ 1- \left ( \frac{\upsilon_{i}-\upsilon_{\star}(t)}{\upsilon \: sin \: i} \right )^{2} \right ]^{1/2} +  c_{2}\left [ 1- \left ( \frac{\upsilon_{i}-\upsilon_{\star}(t)}{\upsilon \: sin \: i} \right )^{2} \right ]. 
\end{multline}

$G(\upsilon_{i},t)$ is the rotational broadening kernel applied over velocities $\upsilon_{i}$ and epochs, $t$, while $\upsilon_{\star}$ is the RV of the star with respect to the solar system. The constants $c_{1}$ and $c_{2}$ are computed as follows:

\begin{equation}\label{eqn:c1}
c_{1} = \frac{2(1-\epsilon)}{\pi \upsilon \: sin \: i (1-\epsilon /3)}
\end{equation}

and

\begin{equation}\label{eqn:c2}
c_{2} = \frac{\epsilon}{2 \upsilon \: sin \: i (1-\epsilon /3)}
\end{equation}

where $\epsilon$ is the linear limb darkening coefficient describing how the center of the stellar or substellar target appears brighter than the edge.

\par As the linear limb darkening law represents a poor approximation of the limb darkening phenomenon for stellar and substellar targets, we conducted a comparison between the rotational broadening kernel in Equation \ref{eqn:RotationalBroadEqn} and one created via higher-order Claret limb darkening. Using \texttt{Scipy}'s Least Squares tool \citep{scipy2020}, we fit Equation \ref{eqn:RotationalBroadEqn} to a LP created with Claret's limb darkening coefficients for a 1500 K, $\log$(g) $= 5.0$, solar metallicity, substellar object \citep{Claret2012}. Both profiles were also convolved with an instrument profile. The resulting profiles agreed within 0.2\%, demonstrating Equation \ref{eqn:RotationalBroadEqn}'s suitability for representing ultracool stellar and substellar disks.

\subsubsection{Gaussian Spots}

\par Surface inhomogeneities can be described by Gaussian distributions centered on an RV value that is a function of the longitudinal viewing angle with respect to the observer. The Gaussian distribution's width is set by the surface inhomogeneity's longitudinal radius. The magnitude of the Gaussian distribution is a product of its area and contrast.  

\par As the target star or substellar object rotates, the Gaussian distribution will transit across the LP. The range of the transit in RV space depends on the inhomogeneity's latitude. As demonstrated in Figure \ref{fig:DevPlotExample}, an equatorial inhomogeneity will transit the entire rotationally-broadened LP while an inhomogeneity at mid-to-polar latitudes will transit an RV band commensurate with its latitude. Conversely, a polar spot on an inclined target may not rotate out-of-view such that while its radial velocity domain is limited by its latitude, its signature will be viewable throughout the entire phase space.

\par We follow the example of \citet{crossfield14}, \citet{karalidi15}, and \citet{karalidi16}, by modeling surface inhomogeneities as circles or ellipses. These inhomogeneities are initially modeled as two-dimensional Gaussian ellipses with the longitudinal and latitudinal widths adjusted for viewing angle by Lambert's Cosine Law,

\begin{equation}\label{eqn:LambertCosineLaw}
I(\theta) = I_{0} \: cos (\theta),
\end{equation}

which states that the observed intensity of an Lambertian object decays according to a cosine function where $\theta$ is the angle of the observer with respect to the normal. Assuming a circular Gaussian spot, the longitudinal ($r_{L}$) and latitudinal ($r_{l}$) radii then become:

\begin{eqnarray}\label{eqn:SpotRadiusLong}
r_{L} = r \: cos(\alpha) \nonumber \\
r_{l} = r \: cos(\beta)
\end{eqnarray}

where $r$ is the radius of the Gaussian spot with units of degrees and $\alpha$ and $\beta$ are the longitudinal and latitudinal viewing angles. These viewing angles are measured from the sub-observer centroid of target to the center of the Gaussian spot. They are computed by first assuming that the target is a sphere with radius equal to unity and the following spherical coordinates,

\begin{eqnarray}\label{eqn:geo2spherical}
r & = & 1 \nonumber \\
\theta & = & 90 - l \nonumber\\
\phi & = & L. 
\end{eqnarray}

The spot's latitude and longitude are then converted into Euler coordinates, $\vec{x} = (x,y,z)$,

\begin{eqnarray}\label{eqn:spherical2euler}
x & = & r \ cos(\phi) \ sin (\theta) \nonumber \\
y & = & r \ sin(\phi) \ sin (\theta) \nonumber \\
z & = & r \ cos (\theta). 
\end{eqnarray}

We define the rotational angle $(\xi)$ as a function of the inclination $(\xi = 90^{\circ}-i)$. The transformation matrix $(T)$ is composed of rotation parameters $(a,b,c,d)$. Computing these rotation parameters, we encounter the components of the unit vector $\vec{k} = (k_x,k_y,k_z)$, but as we will only rotate along the $y$-axis by varying the inclination, $k_x = k_z = 0$ \citep{Shuster1993}.

\begin{eqnarray}\label{eqn:rotationparams}
a & = & cos \big(\frac{\xi}{2}\big) \nonumber \\
b & = & k_x \ sin \big( \frac{\xi}{2} \big) = 0 \nonumber \\
c & = & k_y \ sin \big( \frac{\xi}{2} \big) \nonumber \\
d & = & k_z \ sin \big( \frac{\xi}{2} \big) = 0
\end{eqnarray}

The rotated Euler coordinates $(\vec{x'})$ can now be computed with the following transformation \citep{Shuster1993},

\begin{equation}\label{eqn:EulerRodriguesTrans}
\vec{x'} = T \vec{x}
\end{equation}

where $T$ is defined as follows \citep{Shuster1993}:

\begin{multline}\label{eqn:EulerRodrigues}
T = \\
\left[
\begin{array}{c c c}
\ a^2+b^2-c^2-d^2 & 2(bc-ad) & 2(bd + ac) \\
 2(bc+ad) &  a^2+c^2-b^2-d^2 &   2(cd-ab) \\
 2(bd-ac) & 2(cd + ab) & a^2 + d^2-b^2-c^2 
\end{array} \right] 
\end{multline}

\par The Euler-Rodrigues formula is used to account for the coordinate transformation necessitated by any orbital inclination that is not $90^{\circ}$. The transformation results in new coordinates ($\vec{x'}$) which can then be converted back into the longitudinal and latitudinal viewing angles, $\alpha$ and $\beta$. Using Equation \ref{eqn:SpotRadiusLong}, the longitudinal radius, $r_{L}$ is then converted into RV space for use in the LP; this will be the standard deviation $(\sigma)$ of the Gaussian spot.

\par Using the Euler-Rodrigues formula to compute a spot's position with respect to the observer's viewing angle creates a set of coordinates which, depending on the duration of observations, could include spot positions encompassing an entire revolution. In subsequent analysis, we will compute the longitudinal and latitudinal viewing angles for one complete revolution. Depending on the target's inclination, along with the spot's latitude, the spot may have instances where it is not viewable for periods of time. In the opposite case, for positive inclinations $(0^{\circ} < i < 90^{\circ})$, if the latitude is greater than inclination ($l > i$), spots in the Northern Hemisphere (hemisphere tilted towards the observer) will remain in view for the entire revolution as they circle the North Pole. Each of these cases must be taken into account and spot-induced LP deviations will be set to zero when the spot is not within view of the observer.

\par The magnitude of the Gaussian spot is a function of the spot's area and contrast. The area ($a_{spot}$) is computed as a fraction of the star's unit disk:

\begin{equation}\label{eqn:spotarea}
a_{spot} = \frac{r_{L} \: r_{l}}{(90^{\circ})^2}
\end{equation}

This leads us to the Gaussian distribution for the surface inhomogeneity:

\begin{equation}\label{eqn:GaussianSpot}
p(RV) = \frac{a_{spot} \cdot contrast}{\rm{\sigma}\sqrt{2 \pi}} \ e^{ -\frac{1}{2} \left( \frac{RV-RV_{spot}}{\rm{\sigma}} \right)^2}
\end{equation}

where, 

\begin{equation}
\sigma = r_L \frac{v \ sin \ i}{90^{\circ}}
\end{equation}

and

\begin{equation}
RV_{spot} = \alpha \frac{v \ sin \ i}{90^{\circ}}.
\end{equation}

\par $RV_{spot}$ is the spot centroid RV for each epoch and RV ranges from $-3\sigma$ to $3\sigma$ to capture the continuum well beyond the expected rotationally-broadened signal at $\pm 1\sigma$ for normalization purposes. The term $v \ sin \ i / 90^{\circ}$ is used to convert from longitude to velocity space.

\par Contrast varies in the range $[-1,1]$ with $-1$ being twice as bright as the average surface, $0$ demarcating the average surface temperature, and $1$ denoting a perfectly dark spot with $0 \ \rm{K}$ temperature and no photon emission. It can be seen that a negative contrast will lead to more flux added to an LP centered at the spot's location while positive contrast will lead to a subtraction of flux centered at the spot's location. The spot's maximum achievable magnitude is simply the product of its area and its contrast.

\par
\begin{figure*}
\centering
\includegraphics[width=0.9\textwidth]{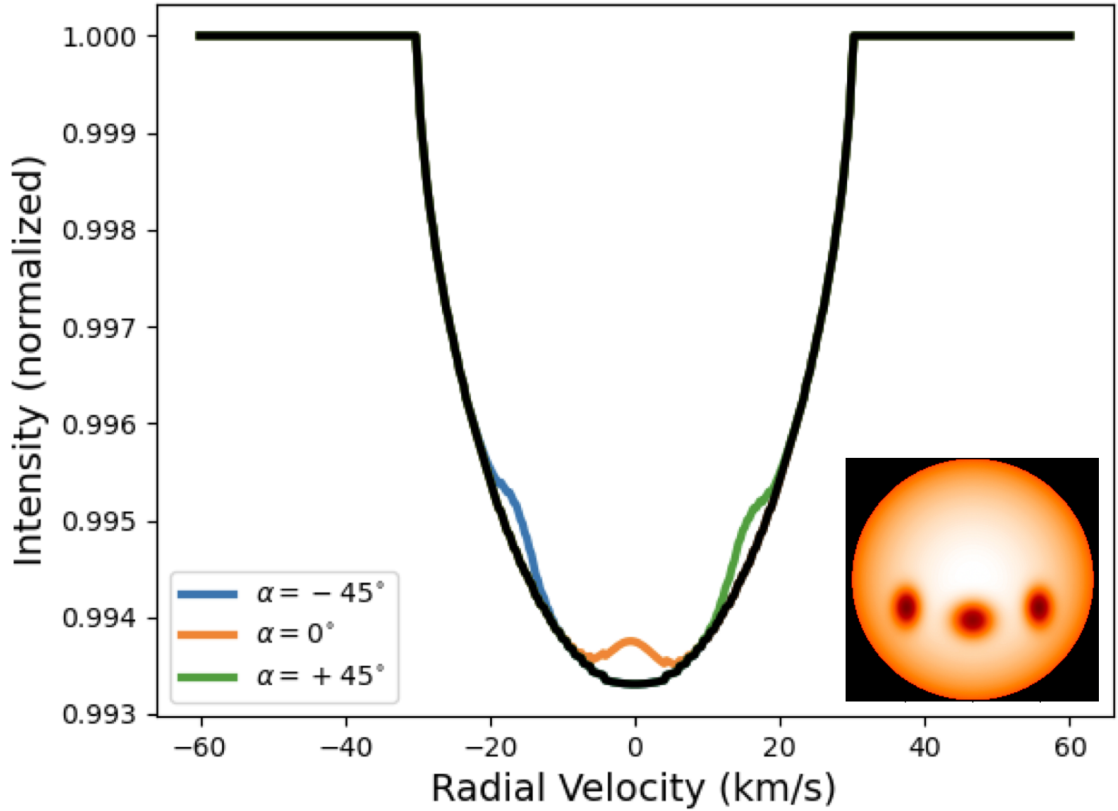}
\caption{\label{fig:analyticalmultispotexample} LPs of spotted target with inclination, $i \: = \: 60^{\circ}$, and viewing longitudes, $\alpha =-45^{\circ},0^{\circ},+45^{\circ}$, with an equatorial latitude, radius of $10^\circ$, and contrast of $0.5$. The black line denotes the unperturbed rotationally broadened LP. (Inset) Corresponding demonstration of a spot at three different longitudes on stellar disk.}
\end{figure*}

 For each epoch, the computed Gaussian spot in the RV domain is then added or subtracted to the rotational broadening LP. Figure \ref{fig:analyticalmultispotexample} demonstrates the analytical method with a star spot moving across a $60^{\circ}$-inclined disk with an equatorial latitude; $-45^{\circ},0^{\circ}$, and $+45^{\circ}$ longitudes; $10^{\circ}$ radius; and $0.5$ contrast. 

\begin{figure*}
\centering
\includegraphics[width=0.9\textwidth]{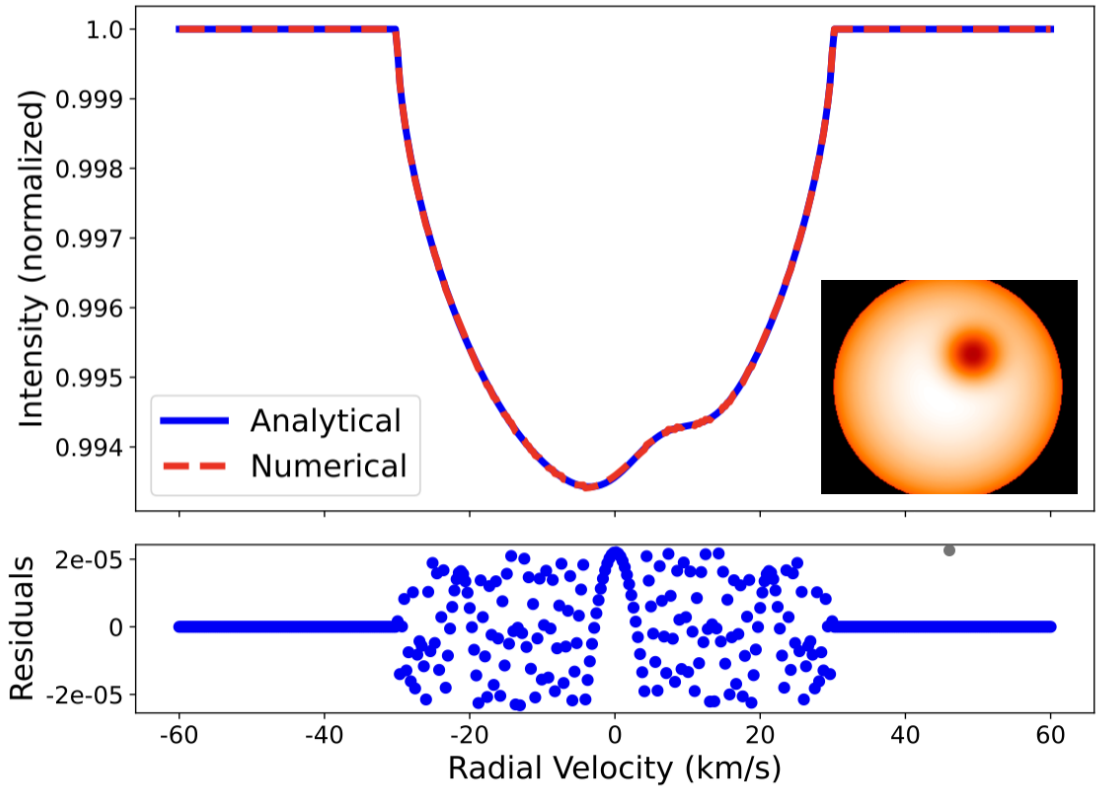}
\caption{\label{fig:NumMapSpotLP}(Top) Analytical and numerical LPs with star spot in the positive RV domain indicating a small blue-shift. (Bottom) Residuals between the analytical and numerical model LP results with maximum amplitude of $2.3 \times10^{-5}$. The small residual values indicate an excellent match between the two different techniques. (Inset) Surface map of stellar surface with spot at $(l = 30^{\circ}, \: L = 20^{\circ})$ with a $15^{\circ}$ radius and contrast of $0.5$ with $i = 85^{\circ}$.}
\end{figure*}

\subsubsection{Analytical Model Photometry}\label{ssec:AnalyticalModel}

\par Although the analytical model was initially created to process observed spectra into LPs, it can be adapted to create photometric light curves integrating the total flux at each phase of the target's rotation. The LSD method incorporates a broad spectral band, so we can assume a featureless and normalized rotational broadening kernel has a summed magnitude approximately the equal to unity at each epoch. Spots with positive contrast (dark spots) will subtract from the total flux at each epoch while spots with negative contrast (bright spots) will add to the total flux.

The produced light curves exhibit a rotationally-broadened shape with a sharp cutoff as a spot enters and exits the stellar limb. This sharp cutoff leads to issues with inferring spot parameters (particularly when applied to real observations such as Luhman 16B in \S \ref{ssec:luhmanmphoto}). To fix this issue, when the analytical code is applied to real-world data, a 1-D Gaussian filter (from the SciPy library) is applied to both analytical and numerical photometric lightcurves \citep{scipy2020}. 

\begin{figure*}
\centering
\includegraphics[width=0.9\textwidth]{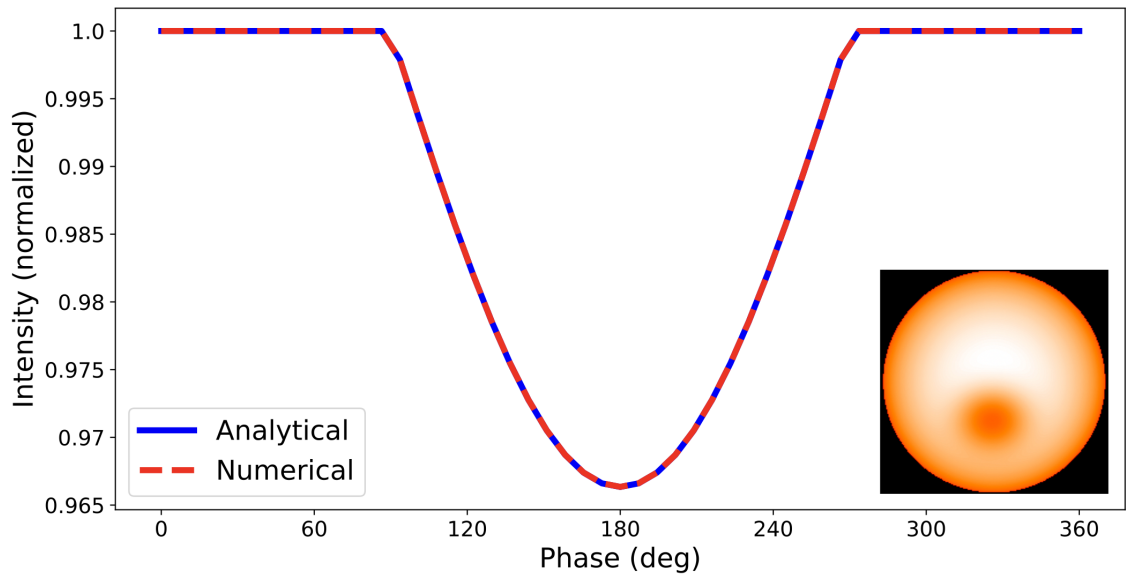}
\caption{\label{fig:photocomp} Photometric light curve comparison for analytical and numerical methods. Light curves include integrated flux summed across the entire disk for each rotational phase angle. A flux reduction of $>3 \%$ due to the modeled spot can be seen centered at $180^{\circ}$ through the rotational period. Computed residuals between the analytical and numerical models are on the ordered of $10^{-6}$. (Inset) Surface map of stellar surface with spot at $(l = 0^{\circ}, \: L = 180^{\circ})$ with a $20^{\circ}$ radius and contrast of $0.3$ with $i = 85^{\circ}$.}
\end{figure*}

\subsection{Numerical Model}\label{ssec:numericalmodel}

While an analytical star spot model has clear advantages in terms of elegance and computational speed, a numerical model for stellar and substellar photospheric surfaces is necessary to verify the validity of the analytical approach. Building a numerical model for the star also provides a convenient option for creating global maps of the star's surface to verify each spot's location as well as their motion (ensuring that the motion computed by the analytical method is indeed physical and passes the so-called `visual check'). 

\par To create the numerical model, the star's surface is divided $n$ times longitudinally and $n/2$ times latitudinally to create $n^{2}/2$ cells. To speed computation, $n$ longitudinal cells are implemented so that each longitudinal cell directly maps to an RV value. For example, a longitude of $+90^{\circ}$ directly maps to an RV of $+ \upsilon \ sin \ i$. Likewise, a longitude of $-90^{\circ}$ maps to an RV of $- \upsilon \ sin \ i$. 

\par It is also necessary to consider the unseen hemisphere. The observer only has a view of one hemisphere, so the flux at any point located at an angular distance greater than $90^{\circ}$ from the sub-observer point ($\alpha = 0^{\circ}, \: \beta = 0^{\circ}$), must be set to zero. 

\par All the points within the $90^{\circ}$ radius are initially set to $1 $ before being multiplied by a linear limb darkening law (to align with the linear limb darkening used in the rotational broadening equation in \S \ref{ssec:AnalyticalSubsection} as well as to match the practice in \cite{crossfield14}). The insets of Figures \ref{fig:DevPlotExample} and \ref{fig:analyticalmultispotexample} demonstrate the numerical code's ability to produce limb-darkened stellar surfaces.

\par Star spots can then be added to the numerical model using a Bi-variate Gaussian distribution with the longitudinal and latitudinal radii (standard deviations) calculated with Equation \ref{eqn:SpotRadiusLong}. As in the analytical technique, the Euler-Rodrigues formula is used to transform the spot's coordinates based on the target's inclination for one entire rotation. The spots are then added or subtracted (bright spots are added while dark spots are subtracted) to the limb darkened image.

\par The numerical code can be implemented to generate a dynamic LP which accounts for the rotation of spots into and out of view as well as the target's inclination. As mentioned above, the target's longitude directly maps to RV in the LP. Figure \ref{fig:NumMapSpotLP} demonstrates the star spot's effect on the LP. Both the numerical and analytical codes lead to a smaller absorption profile in the red-shifted hemisphere which aligns with our expectations that the LP should experience a blue-shift when a surface inhomogeneity is in the Eastern Hemisphere (the hemisphere rotating away from the observer). 

Similar to the analytical model, the numerical code can also be used to generate photometric light curves. This is accomplished by summing up the total flux across the numerical map at each epoch. This will create a value which far exceeds unity, so normalization is necessary. To accomplish this normalization, a featureless numerical map of the star or substellar object is generated using the selected limb darkening law. The sum of flux at each epoch is then divided by this featureless sum of flux to create a normalized value for each measurement. A light curve closely matching the analytical photometry is produced, as seen in Figure \ref{fig:photocomp}. 

\par A similar issue arises with the numerical curve as that seen with the analytical curve, the spot's curve is not smooth. We apply the same solution and use a 1D Gaussian filter to smooth the curve leading to a more physical appearing light curve that also aids in fitting data with Bayesian inference.

\section{Model Validation}\label{sec:validationsection}

\par We validate the analytical model (\S \ref{ssec:AnalyticalSubsection}) against the numerical mode (\S \ref{ssec:numericalmodel}). This is necessary to show that the two models are approximately interchangeable, so it follows the analytical model can be used as a replacement for the numerical method when performing potentially computationally expensive posterior inferences. 

\par For a 1-Spot case, the analytical and numerical models' spectroscopic and photometric results match both qualitatively and quantitatively. Figure \ref{fig:NumMapSpotLP} directly compares LPs generated from the analytical and numerical methods. Residual errors are small on the order of $10^{-5}$. A comparison of the photometric light curves created by the analytical and numerical methods can be seen in Figure \ref{fig:photocomp}. Again, the residual is small with error on the order of $10^{-6}$.

\par Expanding our test cases, an \textit{injection-retrieval} technique is used. Two spots are assigned to a stellar surface in the numerical model with each spot possessing a latitude, longitude, radius, and contrast to create a simulated observation and associated LP. Bayesian inference via nested sampling is then used to retrieve the two spots' locations with our analytical method. 

\par We adopt a similar technique to \citet{karalidi15} where we explore the angular distance, in terms of latitudinal $(\Delta l)$ and longitudinal $(\Delta L)$ displacement, between the two spots. Both spots must be inferred simultaneously using a 2-Spot model as subsequent retrievals with a 1-Spot model will only infer the spot with the largest LP deviation unless restrictive priors are applied. In terms of priors, we assume uninformative priors for the first spot for both longitude ($L \pm 180^{\circ}$) and latitude ($l \pm 90^{\circ}$). For the second spot, we assume either a greater latitude or longitude than the first to overcome the multi-modality of the solution. These tests will be accomplished for varying stellar inclinations with respect to the observer and demonstrate improved results to \cite{karalidi15} in terms of the spatial difference required to resolve to stellar/substellar surface features. 

\subsection{Nested Sampling and Inference}\label{ssec:nestedsampling}

\par We use the open source Python module, \texttt{Dynesty}, to infer spots' locations, size, and brightness using our analytical model as well as assumed priors \citep{speagle20}. For our purposes, performing Bayesian inference on stellar/substellar surfaces using Markov-Chain Monte Carlo (MCMC) methods does not necessarily result in optimal or easy-to-decipher posterior probabilities due to the multi-modal nature of star spots and BD atmospheric features once you move past a 1-Spot model. For this reason, a nested sampling approach was adopted. \texttt{Dynesty} not only functions well with multi-modal data, it also computes the marginal likelihood or, in nested sampling terms, the evidence \citep{speagle20}. 

\par John Skilling developed nested sampling to solve for the evidence by computing the likelihood and prior mass volume through series of \textit{nested} volumes of likelihood \citep{skilling04,skilling06}. Live points are distributed throughout the prior volume and at each iteration the live point with the lowest likelihood is replaced with a new live point with a greater likelihood; the discarded live points (dead points) are then used to compute the evidence and posterior distribution \citep{skilling04,skilling06}. 

\subsection{Spectroscopic Validation}\label{ssec:validationspectral}

\par In this section, we will perform a sequence of validation tests, each resolving two spots with varying angular separation and inclination.

\subsubsection{Test Cases based on Karalidi et al., 2015}

\par To accomplish a systematic exploration of the model's robustness, we adopt a similar approach to \citet{karalidi15} which validated \texttt{Aeolus}, a photometric MCMC-based substellar surface mapping code. \citet{karalidi15} tested their code by varying the angular distance in terms of latitude and longitude for two spots with an approximately infinite SNR. Spot coordinates were input and the model was used to retrieve both spots location. For this work, the numerical model was implemented with input spot locations, sizes, and contrasts to generate simulated observed LPs. We then performed a Bayesian inference using \texttt{Dynesty} to determine the spot locations. 

\par As in \citet{karalidi15}, we assume a larger spot (fixed at the equator at the observer's latitude) with a radius of $9^{\circ}$ and contrast of $0.3$. The smaller spot (with variable location) has a radius of $5^{\circ}$ and contrast of $0.6$. For this validation test, $v \ sin \ i = 50 \ \rm{km}\cdot \rm{s}^{-1}$ was selected based on the rotational range recommended by \citet{vogt87}, and the inclination is nearly edge-on with $i = 85^{\circ}$. 

\par
\begin{figure*}
\centering
\includegraphics[width=1\textwidth]{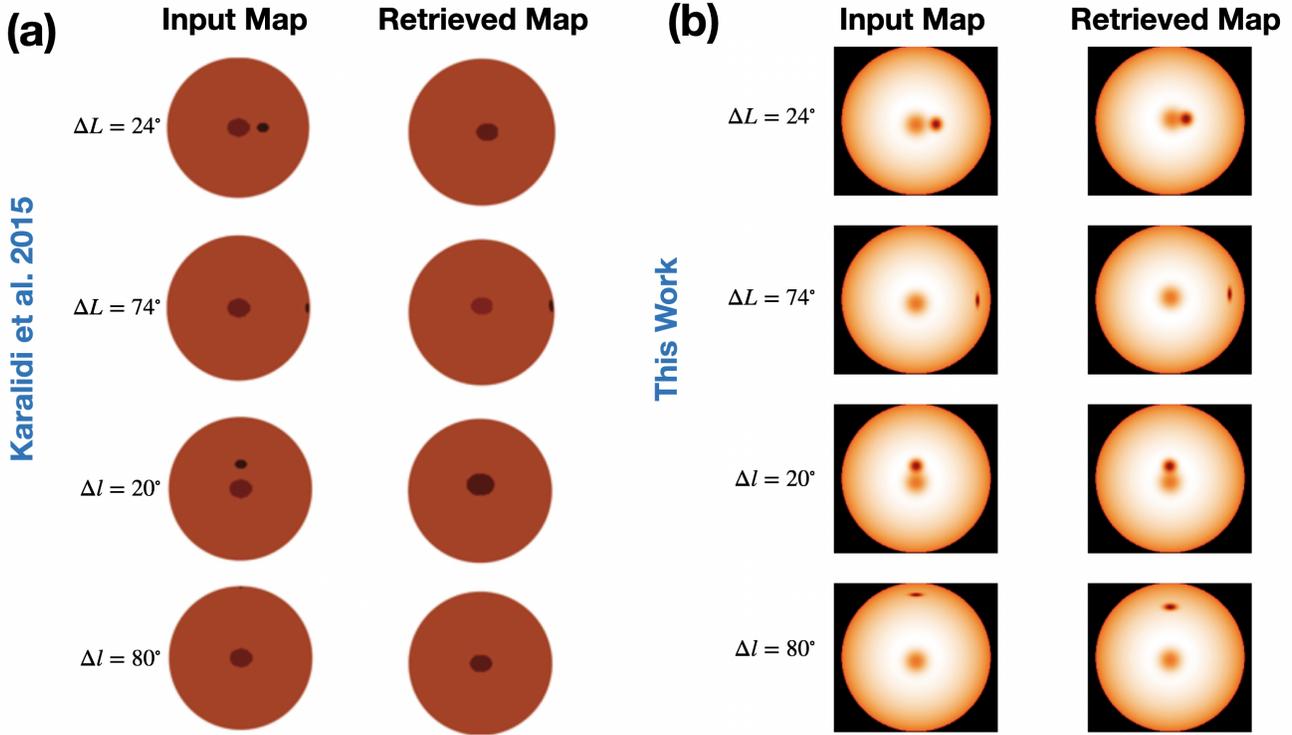}
\caption{\label{fig:KaralidiComp} Spectral \textit{Injection-Retrieval} Comparison. (a) Figure from \citet{karalidi15}. Spots are input and retrieved with longitudinal ($\Delta L$) and latitudinal ($\Delta l$) displacements of $24^{\circ}$, $74^{\circ}$, $20^{\circ}$, and $80^{\circ}$ respectively. Separate spots are not retrieved for the $\Delta L =24^{\circ}$, $\Delta l = 20^{\circ}$, and $\Delta l = 80^{\circ}$ cases. (b) Input and retrieval maps created using numerical and analytical models in this work. For each case, both input spots are retrieved within $1 \sigma$. Simulated LPs were created using the numerical model based on spot locations, and these locations were inferred using nested sampling with the analytical model likelihood implementation.
}
\end{figure*}

 For each case, the input locations are retrieved within $1\sigma$ for both spots. As can be seen in Figure \ref{fig:KaralidiComp}, the analytical model in this work improves on the angular distance that can be resolved: both the smaller angular separations of $24^{\circ}$ longitudinally and $20^{\circ}$ latitudinally can be resolved. 

\par We note that the reader may notice in Figure \ref{fig:KaralidiComp} the angular distances from \citet{karalidi15} do not appear to exactly match the distances from the analytical model from this work. This is an artifact relating to how the maps in \citet{karalidi15} are constructed with two-dimensional circles while the maps generated in this work are effectively three-dimensional due to the Euler-Rodrigues transformation. For example, an equatorial spot traveling $10^\circ$ across a stellar disk will appear to travel less because of the larger radial motion along the line of sight than the tangential motion.

\par Another important difference between the two models is this paper's adoption of Lambert's Cosine Law. In Figure \ref{fig:KaralidiComp}, the effects of the law can be seen in how the spot's shape is modified near the stellar limb to account for the foreshortening of the received flux for a spot seen at high viewing angles.

\par
\begin{figure*}
\centering
\includegraphics[width=0.8\textwidth]{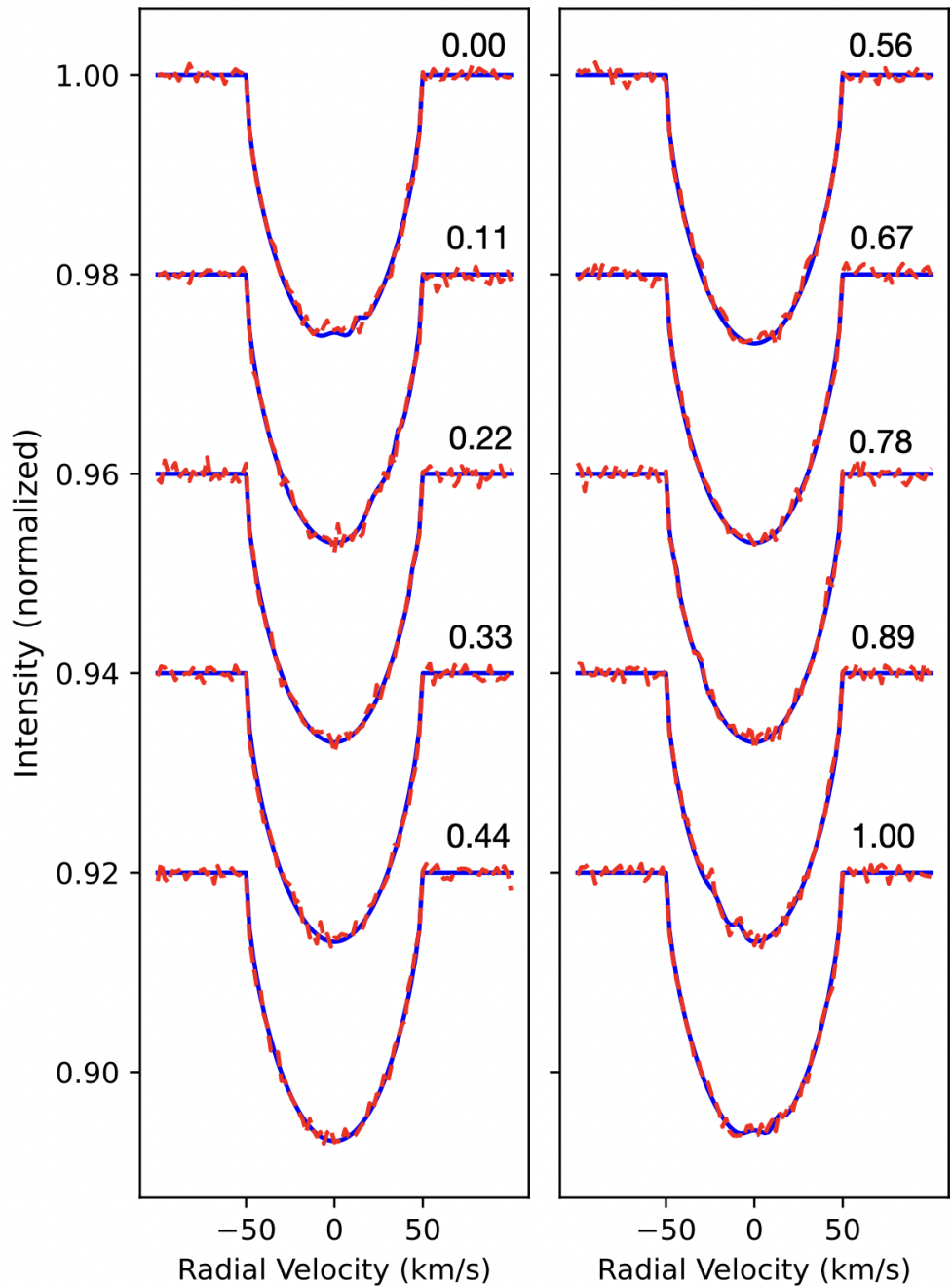}
\caption{\label{fig:LPexample} Example LPs for longitudinally displaced ($\Delta L = 24^{\circ}$) case. Includes 10 epochs, representing one rotation (represented in the figure by phase angle). Dashed red lines represent input numerical spectra with 1\% noise level while the solid blue lines denote the 2-Spot analytical model. LPs are offset by $0.02$ for comparison, but are normalized to unity within the code. The deviations due to surface inhomogeneities can be seen in both the input and retrieved LPs.
}
\end{figure*}

 An example of the input simulated LP data and the retrieved 2-Spot analytical model for the $\Delta L = 24^{\circ}$ case can be seen in Figure \ref{fig:LPexample}. Deviations for both spots can be seen to transit the LP during the early and late phases of the rotation. The 1\% noise level can also be observed in the simulated data.

\subsubsection{Comprehensive Grid Test}\label{ssec:gridtest}

\par Expanding on the comparison shown in Figure \ref{fig:KaralidiComp}, a more exhaustive parameter space validation was conducted with longitude and latitude displacements varying from $0^{\circ}$ and $60^{\circ}$. A realistic noise level of $1 \%$  was selected and radii and contrast were set to $20^{\circ}$ and $0.3$, respectively, to reflect previous BD Doppler imaging results for cloud system radii from \citet{crossfield14} and \citet{luger21a} as well as typical spot temperatures observed in the Sun's photosphere. Inclinations of $30^{\circ}$ and $60^{\circ}$ were selected to explore the code's ability to resolve spots and break degeneracies (seen at edge-on inclinations) at different viewing angles. As above, the numerical model was used to generate simulated observed LPs and nested sampling was used to infer spot locations using the analytical model. 

\par For each trial, we reviewed the results in the form of corner plots and judged each trial as a \textit{pass}, \textit{blend}, or \textit{fail}. A \textit{pass} would be awarded if the inferred locations for both spots were within $1\sigma$ of the true values with 1 distinct peak; a \textit{blend} would be awarded if the $1\sigma$ range for the two spots overlapped in either longitude or latitude; and a \textit{fail} would awarded if either condition above was met. All cases passed the test for a grid with $\Delta L$ and $\Delta L$ from $0^{\circ}$ to $60^{\circ}$ with increments of $15^{\circ}$. For the case of an irregularly-shaped spot, the analytical model would likely interpret this as either multiple overlapping individual inhomogeneities or one Gaussian spot with a reduced contrast (due to the portion of non-contrasted background being incorporated into the larger spot).

\subsection{Photometric Validation}\label{ssec:validationphoto}

\par To truly replicate \citet{karalidi15}'s results, we needed to compare our photometric model to that shown in their work. To this end, we performed the same analysis as shown in \S \ref{ssec:validationspectral} only with photometric data. 

\par Again, we set the rotational speed  to $v \ sin \ i$ to $50 \ \ \rm{km}\cdot \rm{s}^{-1}$ and the viewing angle to nearly edge-on, $i = 85^{\circ}$. The primary and secondary spots were at the same locations with identical spot radii and contrasts as outlined in \S \ref{ssec:validationspectral}. The numerical model was used to generate a photometric light curve as described in \S \ref{ssec:numericalmodel} while the analytical model was used to compute the likelihood within a nested sampling Bayesian inference. Both the analytical and numerical photometric light curves were smoothed using a Gaussian filter as described in \S \ref{ssec:AnalyticalSubsection} and \S \ref{ssec:numericalmodel}. 

\par The results are very similar to Figure \ref{fig:KaralidiComp} with improved spot resolution versus that seen in \cite{karalidi15}. Within the context of this sample case, the analytical code resolves both spots location for both spectroscopic and photometric data. 

\par As in \S \ref{ssec:validationspectral}, we conducted an exhaustive exploration for varying longitude and latitude displacements for $i = 30^{\circ}$ and $60^{\circ}$. The spot size and contrast as well as the noise level ($1 \%$) were identical to \S \ref{ssec:gridtest}. Just as in that section, the photometric analytical model successfully retrieved every input scenario. 

\section{Application to Luhman 16B}\label{sec:luhmansection}

\par With the analytical Doppler imaging method validated versus an numerical model, the next step is to perform an external validation with real-world observational data. Similar to \citet{luger21a}, we choose to validate our technique against Luhman 16B (WISE J104915.57-531906B). 

\subsection{Luhman 16AB System}\label{ssec:L16ABintro}

\par Luhman 16AB are two of the solar system's nearest neighbors at $1.9955 \pm 0.0004 \ \rm{pc}$ which also makes the two binary components the nearest BDs \citep{bedin17}. Luhman 16AB was originally discovered by \citet{luhman13} using images from the Digitized Sky Survey, the Two Micron All-Sky Survey (2MASS), and the Deep Near-Infrared Survey of the Southern Sky (DENIS). Luhman 16AB is comprised of Luhman 16A, a $34.2_{-1.1}^{+1.3} \ M_J$ primary with an L7.5 spectral class, and Luhman 16B, a $27.9_{-1.1}^{+1.0} M_J$ secondary with a measured spectral class T0.5 \citep{burgasser13,kniazev13,garcia17,ammons&garcia19}. 

\par Similar to many BDs at the L/T transition, Luhman 16B demonstrates high photometric and spectroscopic variability \citep{gillon13,biller13,burgasser14,Buenzli15b,Buenzli15a}. Although the primary, Luhman 16A, is also considered variable, it is to a smaller degree than the secondary, Luhman 16B \citep{biller13,burgasser14,crossfield14,Buenzli15b}. Several studies over the last decade have investigated Luhman 16B's variability and periodicity. \citet{gillon13} deduced an $~11\%$ peak-to-peak near-infrared ($0.9 \ \rm{\mu m}$) variability and $4.87 \pm 0.01 \ \rm{hr}$ period over a 12 night campaign. \citet{burgasser14} estimated a similar $13.5 \%$ near-infrared ($1.25 \ \rm{\mu m}$) variability. \citet{apai21} used TESS to observe 100 rotations of the binary system, finding maximum peak-to-peak variability to be approximately $4 \%$ with longer period, $10 \%$ variability seen over $100 \ \rm{hr}$ timelines. \citet{apai21} attributed the strongest periodicity signal ($5.28 \ \rm{hrs}$) along with a secondary periodicity ($2.5 \ \rm{hrs}$) to the combined effects of differential rotation and planetary waves in Luhman 16B's atmosphere. In line with the above results, \citet{heinze21} observed a $5 \ \rm{hr}$ period and variability in the red continuum changing from $7 \%$ to $13 \%$ on consecutive nights, providing a relatively consistent measured period and high-variability.

\subsection{Previous Global Maps of Luhman 16B}\label{ssec:previousL16B maps}

\par Global surface maps of Luhman 16B's surface inhomogeneities have been generated from both spectroscopic and photometric observations. \citet{crossfield14} inferred spot latitude, contrast, and size based on a 1-Spot model. They also created the first surface map of the BD by applying the Maximum Entropy method outlined in \citet{vogt87}. 

\par More recently, \citet{luger21a} reprocessed the \citet{crossfield14} data and used the open-source Python code \texttt{Starry} \citep{Luger2019} to generate a global surface map. The map created by \cite{luger21a} is qualitatively similar to \citet{crossfield14} with both studies deducing an dark equatorial region and a bright polar spot. 

\par Using photometric data from the HST, \citet{karalidi16} generated three and four spot surface maps for Luhman 16A and B. For Luhman 16B, a four spot model with temperature (contrast) as a free parameter was qualitatively different from the maps created by \citet{crossfield14} and \cite{luger21a}, but still contained near equatorial spots and brighter spots in the polar region \citep{karalidi16}. In the following sub-sections, we will use archival data from both of these studies to produce Luhman 16B surface maps with the analytical method described in this paper.

\subsection{Preparing Spectroscopic Data}\label{ssec:SpectralDataPrep}

\par To infer a global map of Luhman 16B's surface, we follow a similar process as that used by \citet{crossfield14} with the same VLT/CRIRES data collected on 2013 May 5. \citet{crossfield14} initially obtained 56 spectra ($2.288-2.345 \ \rm{\mu m}$) with $300 \ \rm{s}$ exposure times and combined these into groups of four to create high SNR LPs via the LSD method. We obtained the Luhman 16B data via open online archival and used the reprocessed template and observed spectra from \citet{luger21a} to similarly create LPs via the process outlined in \S \ref{ssec:LSDsection}. Referencing \S \ref{ssec:LSDsection}, the template spectrum created by \citet{crossfield14} and reprocessed by \citet{luger21a} is used to create a line mask matrix ($M$). Equation \ref{eqn:LSDeqn} is then used to compute the LP for each epoch. 

\par An important point to note is the retrieved latitude's sensitivity to factors such as LP element number ($n$) and the regularization parameter ($\Lambda$). We selected $n = 93$ LP elements to ensure a $\Delta RV \approx 1-2 \ \ \rm{km}\cdot \rm{s}^{-1}$ while also sampling the RV range of $\pm 3\sigma$. $\Lambda = 500$ was qualitatively selected to attenuate noise in the retrieved LP while preserving the observed signal. Varying these two parameters can vary the retrieved latitude, but all resulting variations of retrieved latitudes are consistent with \citet{crossfield14}.

\subsection{Spectroscopic Results}\label{ssec:luhmanspectral}

\begin{figure*}
\centering
\includegraphics[width=0.8\textwidth]{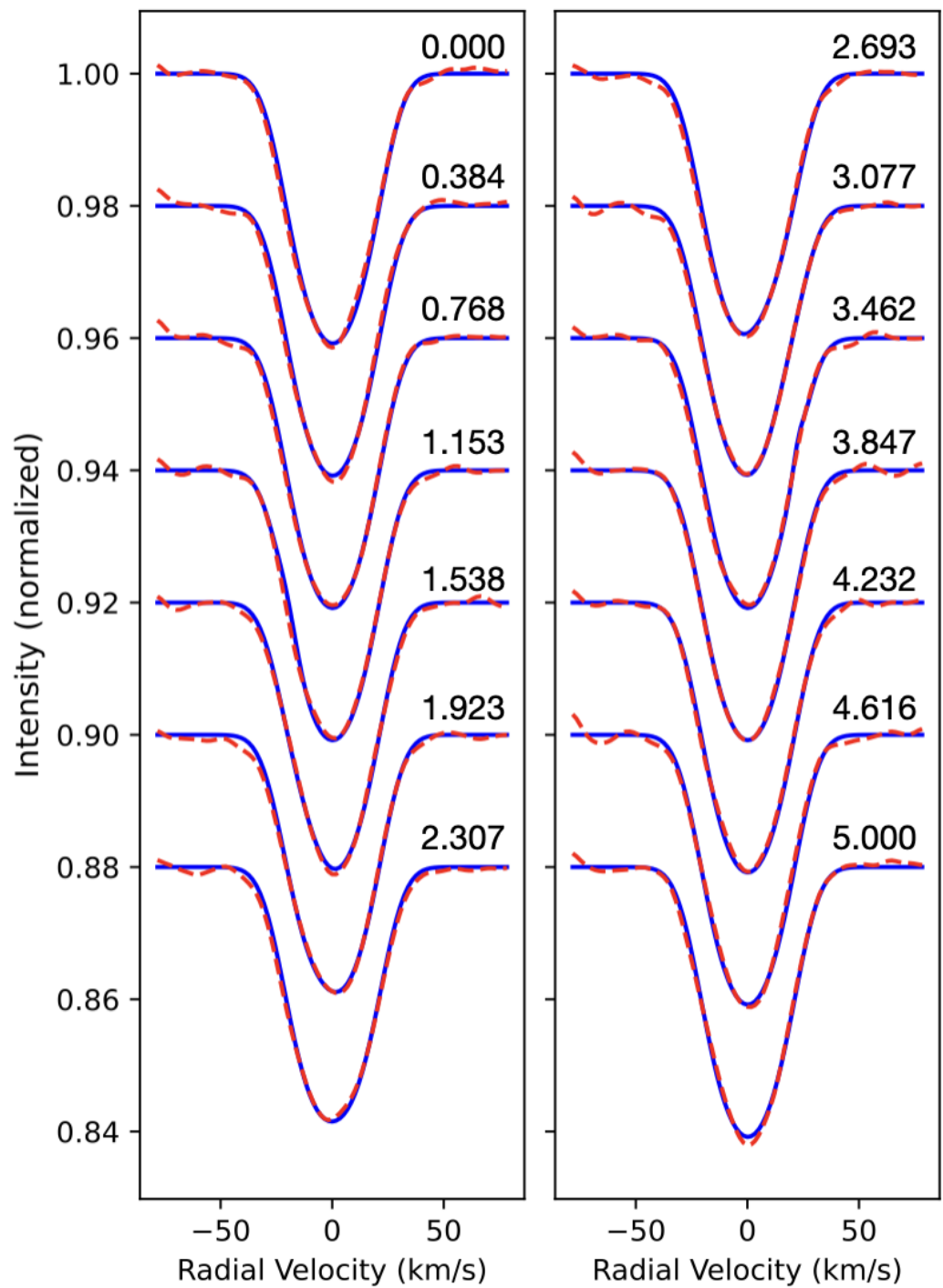}
\caption{\label{fig:LineProfiles} LPs for 14 epochs, representing one rotation of Luhman 16B. Dashed red lines represent observed spectra from \cite{crossfield14} while the solid blue lines denote the 1-Spot analytical model developed in this paper. Each epoch is accompanied by the appropriate time stamp in hours with the period being $\approx \ 5 \ hrs$. LPs are offset by $0.02$ for comparison, but are normalized to unity within the code. The mid-phase LP deviations are difficult to detect visually due to the dominant spot's large radius and low contrast creating shallow flux variations; thereby, necessitating computational methods to retrieve. Note the analytical model has been convolved with a Gaussian instrument profile to account for the spectral resolution.
}
\end{figure*}

The resulting LPs can be seen in Figure \ref{fig:LineProfiles} along with the line fit from our 1-Spot analytical model. Due to the retrieved spot's radius and relatively low contrast (compared to the validation examples), the affected region of the LP is broad and the subtle shifts in flux are difficult to detect by eye, necessitating a computational Bayesian method. Using nested sampling, we inferred spot longitude, latitude, radius, and contrast by comparing the LPs generated from observed spectra to our analytical model. As in \citet{crossfield14} and \citet{luger21a}, we set inclination to $i = 70^{\circ}$ and $v \ sin \ i = 26.1 \ \ \rm{km}\cdot \rm{s}^{-1}$. We manually varied the linear limb darkening coefficient ($\epsilon$) and the number of LP elements ($n$) to fit the observed data. We convolved the LP with a Gaussian profile to account for instrument and other sources of line broadening resulting in the modeled LPs in Figure \ref{fig:LineProfiles}.

\par
\begin{figure*}
\centering
\includegraphics[width=1\textwidth]{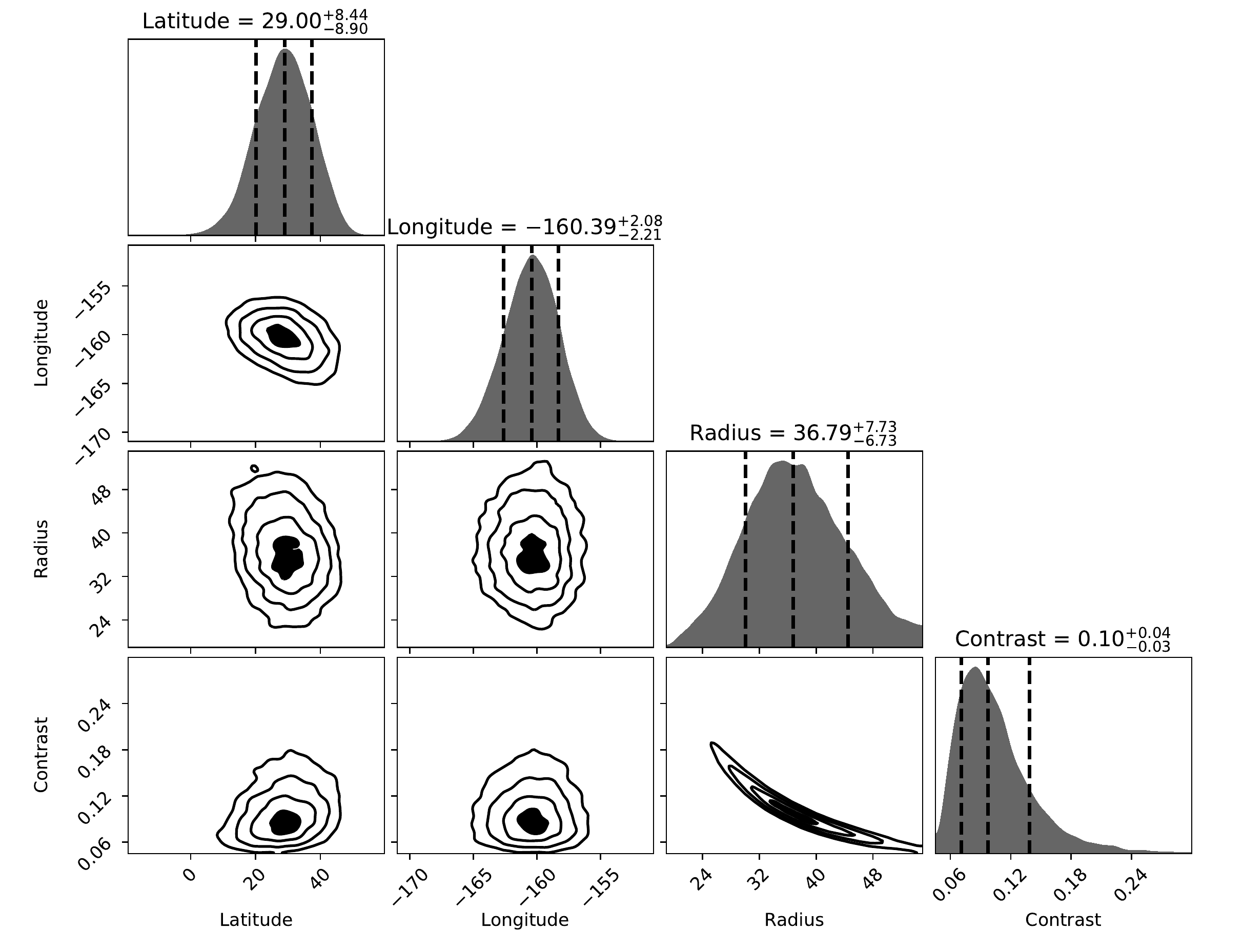}
\caption{\label{fig:Corner1Spot} 1-Spot (Spectroscopic) corner plot. The mean inferred values along with $1\sigma$ quantiles displayed on the top of each column. Contours denote $0.5,1,1.5,$ and $2\sigma$ regions. Created with \texttt{Dynesty} \citep{speagle20}.
}
\end{figure*}

 For the 1-Spot model, dynamic nested sampling resulted in an equatorial dark spot  (latitude $= 29^{\circ}$) with radius $= 37^{\circ}$  and contrast $=0.10$ which can be seen in Figure \ref{fig:Corner1Spot}. These results can be compared to \cite{crossfield14}, \cite{karalidi16}, and the photometric results in Table \ref{tab:ResultsSummary}. Our results match well with the values given in \citet{crossfield14} which inferred a latitude $\leq 31^{\circ}$, radius = $33^{\circ} \pm 7^{\circ}$, and contrast = $0.12 \pm 0.03$. Longitude was not included in \citet{crossfield14} as it is dependent on the beginning epoch of observation.

\par
\begin{figure*}
\centering
\includegraphics[width=1\textwidth]{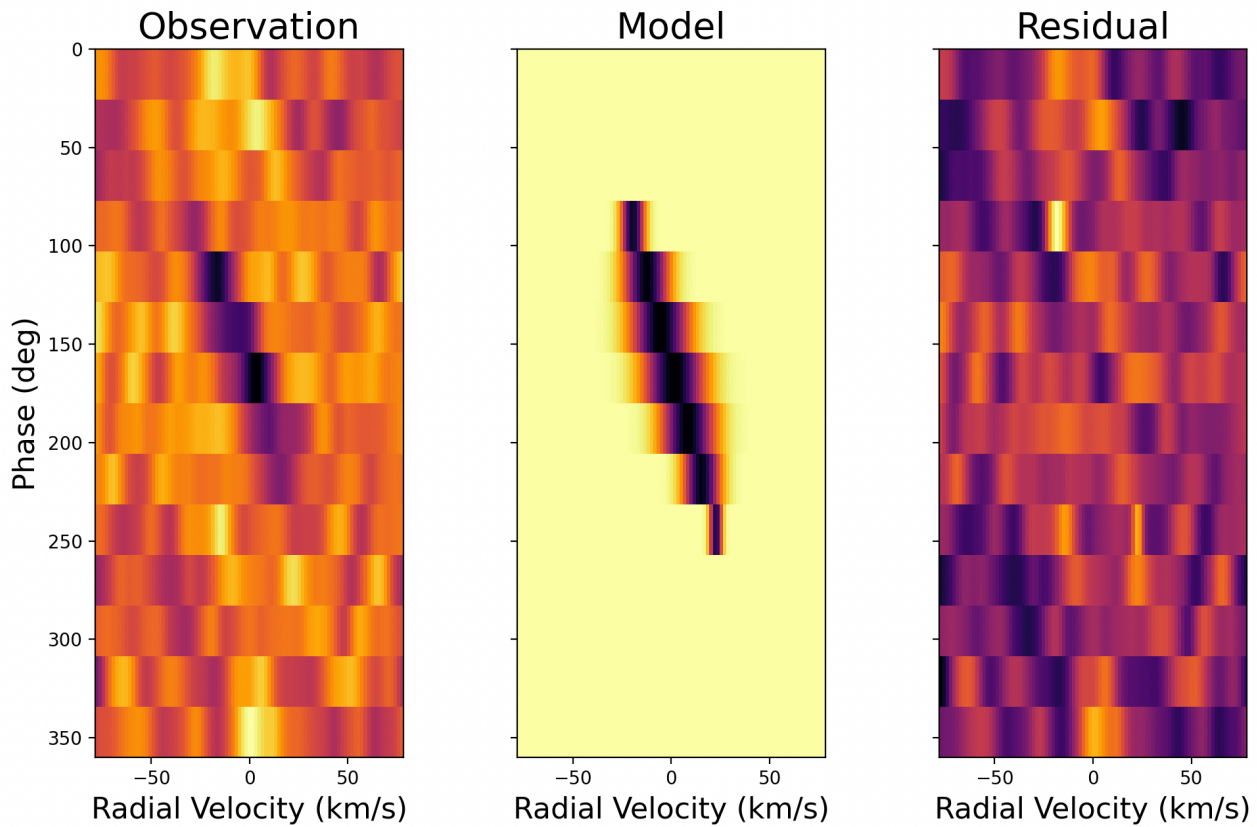}
\caption{\label{fig:DevPlot1Spot} Luhman 16B LP deviations. (Left) Deviations from \cite{crossfield14} observed mean LP. (Middle) Modeled deviations for inferred equatorial dark spot. (Right) Residuals computed by subtracting modeled LP deviations from observed LP deviations.
}
\end{figure*}

 The 1-Spot model reproduces \citet{crossfield14}'s MCMC results and shows them to be a reasonable representation of Luhman 16B's dominant feature, the equatorial dark spot (Figure \ref{fig:DevPlot1Spot}). For models with a higher number of spots, the spectroscopic analytical method did not converge on a likelihood, leading us to believe the 1-Spot model best describes the \cite{crossfield14} data.

\subsection{Photometric Results}\label{ssec:luhmanmphoto}

\par Luhman 16A and B photometric data was collected on 2013 November 8 by the Hubble Space Telescope (HST)/Wide-Field Camera (WFC) with the G141 grism which operates between 1.1 and 1.66 $\mu m$. The data was originally collected, processed, and reported on by \citet{Buenzli15a}. \citet{karalidi16} used those observations, as well as those collected by \citet{Buenzli15b} with the G102 grism (0.8 to 1.15 $\rm{\mu m}$) on 23 November 2014, to construct three and four spot surface maps using the \texttt{Aeolus} code developed in \citet{karalidi15}. 

\par Here we will seek to infer surface spots from the \citet{Buenzli15a} data using the analytical method described in \S \ref{ssec:AnalyticalSubsection}. The $J$ and $H$ near-infrared bands from \citet{Buenzli15a} result in very similar lightcurves, so for this paper these bands are averaged. 

\par
\begin{figure*}
\centering
\includegraphics[width=1\textwidth]{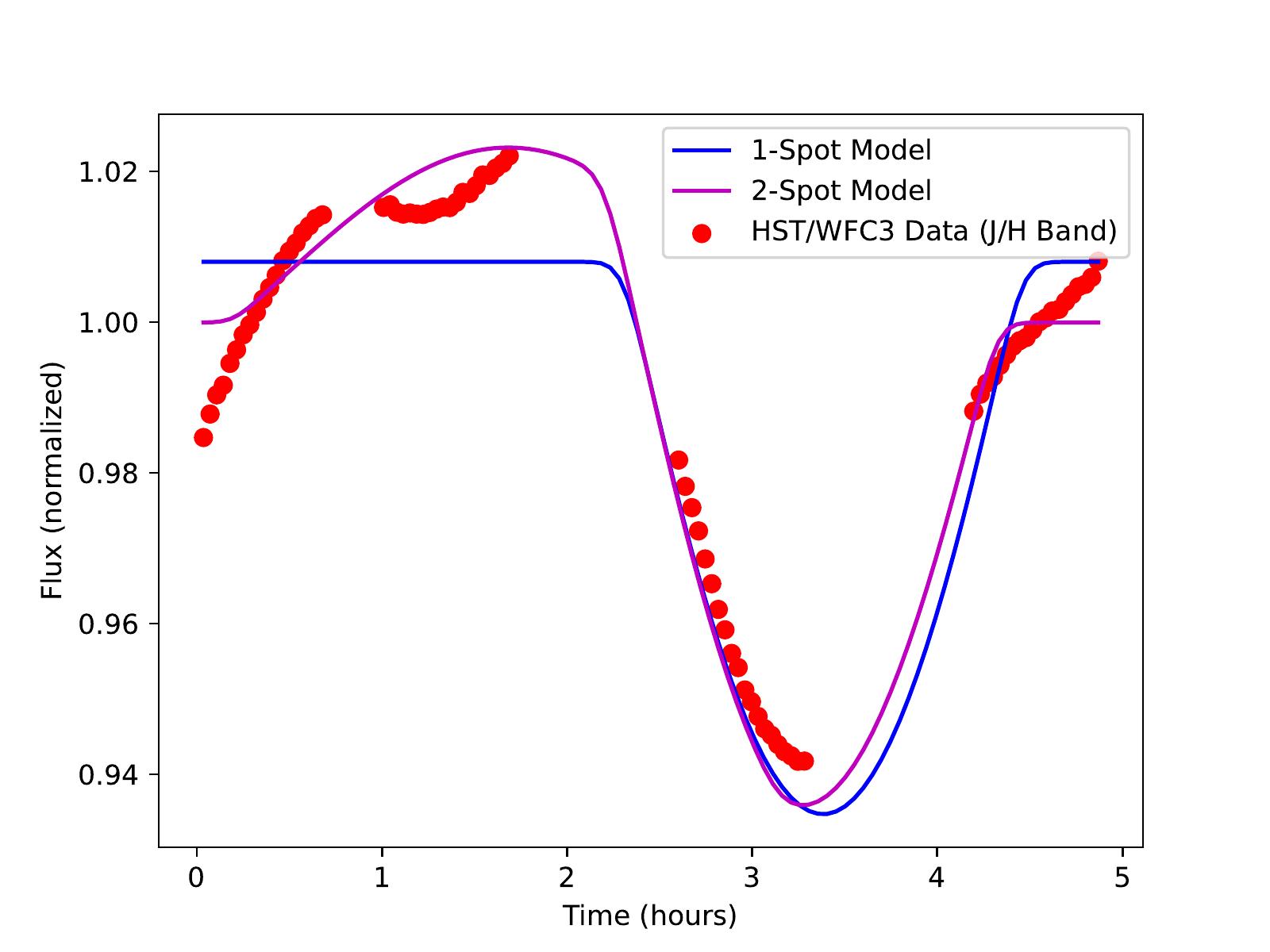}
\caption{\label{fig:L16Blightcurves} Photometric light curves from \cite{Buenzli15a} along with 1 and 2-Spot analytical models. The 1-Spot model includes an equatorial dark spot while the 2-Spot model includes both an equatorial dark spot as well as a mid-latitude bright spot. Based on the results of \citet{karalidi16}, the unmodeled trends in the light curve are potentially due to additional bright and dark spots combined with Luhman 16B's observed short-term timescale variability. 
}
\end{figure*}

 In Figure \ref{fig:L16Blightcurves}, the \citet{Buenzli15a} $J$ and $H$ band data, along with our 1 and 2-Spot photometric models are plotted. The 2-Spot light curve model more fully represents Luhman 16B's photometric behavior. Dynamic nested sampling results in a model with improved evidence ($\Delta \ln \mathcal{Z} = 13.8$) for the 2-Spot model over the 1-Spot model. This is strong evidence that the 2-Spot model better represents the data \citep{Kass&Raftery1995}. A summary of the photometric results can be found in Table \ref{tab:ResultsSummary}.

\par Further increasing the number of spots in our model leads to lower comparative evidence ($\Delta \ln \mathcal{Z}$), so we conclude the strongest evidence exists for the 2-Spot model. The equatorial dark spot and mid-latitude bright spot inferred by the 2-Spot photometric model echo the results seen in \citet{karalidi16}'s photometric results as well as the Doppler imaging maps in \citet{crossfield14} and \citet{luger21a}. 

\par A summary of both spectroscopic and photometric results, as well as comparisons to \citet{crossfield14} and \citet{karalidi16}, can be seen in Table \ref{tab:ResultsSummary}. Both the spectroscopic and photometric results reasonably match those obtained in \citet{crossfield14} and \citet{karalidi16}, respectively. Even though the results from \citet{karalidi16} rely on fixed bright spots (negative contrast for this paper's nomenclature), our 1-Spot and 2-Spot analytical models still infer approximately the average spot location and size of \citet{karalidi16}'s $J$ and $H$ bands. This  matches our expectations as we averaged the $J$ and $H$ before performing Bayesian inference via nested sampling. Furthermore, our 2-Spot photometric model resulted in a Bayesian Information Criterion (BIC) value of 41.4, matching the $J$ band 2-Spot model value achieved in \citet{karalidi16}.

\begin{table*}[t]
\caption{\label{tab:ResultsSummary} Model Comparison Summary (* Denotes where $+360^{\circ}$ has been added to negative longitudes to more easily compare to external results where longitudinal phase goes from $0^{\circ}$ to $360^{\circ}$  versus $-180^{\circ}$ to $180^{\circ}$)}
\centering
    \begin{tabular}{l l c c c c }
    \hline
    \hline
    Model & Mode & Latitude & Longitude & Radius & Contrast \\ 
    \hline
    \cite{crossfield14} & Spectroscopic & $\leq 31^{\circ}$ & N/A & $33^{\circ} \pm 7^{\circ}$ & $0.12 \pm 0.03$ \\ 
    & & & & &  \\
    \hline
    1-Spot & Spectroscopic & $29.00^{\circ^{+8.44^{\circ}}}_{-8.90^{\circ}}$ & $-160.39^{\circ^{+2.08^{\circ}}}_{-2.21^{\circ}}$ & $36.79^{\circ^{+7.73^{\circ}}}_{-6.73^{\circ}}$ & $0.10^{+0.04}_{-0.03}$ \\
    & & & & &  \\
    \hline
    \cite{karalidi16} & Photometric $(J)$ & $-20^{\circ} \pm 12^{\circ}$ & $113.47^{\circ} \pm 6.46^{\circ}$ & $38.89^{\circ} \pm 0.8^{\circ}$ & $-0.180 \pm 0.013$  \\
    & (Fixed Bright Spots)  & $45^{\circ} \pm 12^{\circ}$ & $186.08^{\circ} \pm 3.65^{\circ}$ & $24.45^{\circ} \pm 1.46^{\circ}$ & $-0.180 \pm 0.013$  \\
    &  & $72^{\circ} \pm 10^{\circ}$ & $283^{\circ} \pm 13^{\circ}$ & $35.54^{\circ} \pm 1.67^{\circ}$ & $-0.180 \pm 0.013$ \\
    \hline
    \cite{karalidi16} & Photometric $(H)$ & $-18^{\circ} \pm 8^{\circ}$ & $107.27^{\circ} \pm 5.48^{\circ}$ & $32.42^{\circ} \pm 0.8^{\circ}$ & $-0.180 \pm 0.013$ \\
    & (Fixed Bright Spots)  & $51^{\circ} \pm 20^{\circ}$ & $190.91^{\circ} \pm 4.83^{\circ}$ & $25.02^{\circ} \pm 1.13^{\circ}$ & $-0.180 \pm 0.013$  \\
    &  & $74^{\circ} \pm 6^{\circ}$ & $288.6^{\circ} \pm 9.4^{\circ}$ & $36.30^{\circ} \pm 1.18^{\circ}$ & $-0.180 \pm 0.013$ \\
    \hline
    1-Spot & Photometric $(J/H)$ & $-23.53^{\circ^{+15.72^{\circ}}}_{-9.89^{\circ}}$ & $110.08^{\circ^{+3.30^{\circ}}}_{-2.79^{\circ}}$ & $37.58^{\circ^{+4.86^{\circ}}}_{-5.37^{\circ}}$ & $0.25^{+0.07}_{-0.06}$ \\
    & & & & & \\
    \hline
    2-Spot & Photometric $(J/H)$ & $-27.20^{\circ^{+18.22^{\circ}}}_{-10.86^{\circ}}$ & $119.30^{\circ^{+5.92^{\circ}}}_{-5.22^{\circ}}$ & $34.86^{\circ^{+3.59^{\circ}}}_{-4.85^{\circ}}$ & $0.28^{+0.05}_{-0.06}$ \\
    &  & $47.19^{\circ^{+24.41^{\circ}}}_{-32.52^{\circ}}$ & $^{*}231.68^{\circ^{+15.54^{\circ}}}_{-20.44^{\circ}}$ & $28.20^{\circ^{+7.64^{\circ}}}_{-7.44^{\circ}}$ & $-0.14^{+0.06}_{-0.09}$ \\
    & & & & & \\
    \hline
    \hline
    \end{tabular}
\label{tab:datasets}
\end{table*}

\section{Discussion}\label{sec:discussion}

\subsection{Persistent Atmospheric Features}\label{ssec:discussion_persistentfeautres}

\par Multiple data sets at different wavelengths have explored the surface of Luhman 16B. From these studies, evidence has emerged that the BD possesses a persistent dark equatorial spot and potentially a bright mid-to-polar latitude spot.

\par Persistent atmospheric features for Luhman 16B can be seen in the results of this paper as well as \citet{crossfield14}, \citet{luger21a}, and \citet{karalidi16}. As discussed above, \cite{crossfield14} inferred a dark equatorial spot using MCMC. Using the Maximum Entropy method, \cite{crossfield14} also created a global map which included a dark equatorial spot as well as a bright polar feature. \citet{luger21a} re-processed the same spectroscopic data and inferred a similar map with those same features. \citet{karalidi16} identified the dark feature and labeled it the Possible Persistent Cloud Structure (PPCS-1) which may be similar to Jupiter's Great Red Spot. Not only did it appear in the HST $J$ and $H$ bands as well as a G102 grism light curve, it also appears in data presented in \citet{gillon13} which used the $I + z$ TRAPPIST filter \citep{karalidi16}.

\par These features may be consistent with \cite{apai21}, in which Luhman 16B's planetary waves, potentially in the form of zonal jets, were demonstrated over 100 hours of TESS observations. It is possible the persistent dark spot forms in such a zonal jet at Luhman 16B's equator while mid-to-polar spots form within higher latitude bands. \citet{apai21} also detected what they thought to be a long period signal from a notional polar vortex on Luhman 16A. The existence of such persistent features could help to shed light on the overall atmospheric structure and dynamics of Luhman 16B and similar ultracool objects.

\par While there is evidence for persistent features, variability remains a recurring theme in observations of brown dwarfs. L/T transition variability is likely caused by multiple atmospheric layers and clouds with disparate opacities. The variability of Luhman 16B \citep{gillon13,crossfield14, Buenzli15b, Buenzli15a} may also arise from the different observational techniques and wavelengths considered. Not only did these studies observe Luhman 16B at different times, they are also probed different pressure levels and chemical species. The rich inhomogeneous data sets can complicate the picture when attempting to quantify the true variability of the object. However, they also offer an opportunity to understand the BD's three-dimensional atmospheric dynamics.

\subsection{Doppler Imaging as Model Validation}\label{ssec:discussion/modelvalidation}

\par Over the past decade, there has been extensive work in modeling atmospheric dynamics in lightly irradiated BD atmospheres through general circulation models (GCMs). Several drivers for robust circulation in the stratified overlying atmosphere of these BDs have been explored in the literature including convection driving the stratified upper layer and cloud-radiative feedback. \cite{showman20} and \cite{Zhang2020} both include comprehensive reviews of these efforts.

\par Convection from the interior perturbing the overlying stratified atmospheric layer has been a mechanism identified as a potential primary driver of BD circulation \citep{Freytag2010,showman&kaspi13}. The strength of these perturbations, along with the object's radiative timescale and drag at the radiative-convective boundary (RCB), appear to control the latitudinal extent and strength of jet stream formation \citep{zhang14,showman19,tan22}. 

\par Cloud-radiative feedback has also been demonstrated as a plausible mechanism for driving atmospheric circulation. One-dimensional feedback between clouds, radiation, and atmospheric mixing is explored in \citet{Tan&Showman19}. More recently, modelers have focused on three-dimensional feedback models in which clouds-radiative dynamics induce active atmospheric circulation thereby perpetuating cloud patchiness \citep{tan21a,tan21b,Lefevre2022}.

\par Placing Luhman 16B's large equatorial dark spot and mid-latitude/polar bright spots within the context of a model such as that discussed in \citet{zhang14} offers a means to connect Doppler imaging to atmospheric dynamics. In that work, the atmospheric structure is essentially a product of forcing from the convective interior and the radiative timescale of the overlying stratified atmosphere. Strong convective forcing and weak radiation lead to a large equatorial jets with mid-latitude to polar vortices and turbulence. It is possible that the equatorial dark spots are persistent clouds or storm systems within a jet stream band. The bright spots may be regions of lessened opacity and cloud thickness created by turbulent features in which we can observe deeper layers of the BD's atmosphere. 

\par Combining BD atmospheric circulation models with Doppler imaging techniques could potentially lead to both more accurate surface maps as well as validation for the development and selection of GCMs. Follow-on observations of Luhman 16B and other ultracool targets may result in more accurate atmospheric models.

\subsection{Doppler Imaging and Chemical Disequilibrium}\label{ssec:discusion/disequilibrium}

\par Doppler imaging focused on specific chemical species' spectral lines could help to further illuminate the mechanisms at play in BD and gas giant exoplanet atmospheres. Tracing specific species such as carbon monoxide (CO), methane (CH$_4$), ammonia (NH$_3$), and molecular nitrogen (N$_2$), provides insight into atmospheric circulation. At temperatures above the L/T transition, carbon's chemical equilibrium state transitions from favoring CO (above $1470 \ \rm{K}$) to CH$_4$ (below $1470 \ \rm{K}$) \citep{Fegley&Lodders1996}. A similar chemical equilibrium transition occurs for nitrogen. Below $630 \ \rm{K}$, NH$_3$ should be the dominant species while N$_2$ is favored by chemical equilibrium above this value \citep{Fegley&Lodders1996}.

\par Chemical disequilibrium in ultracool atmospheres provides evidence for robust circulation. The expectation for BD spectra at temperatures at or below the L/T transition is to favor higher relative abundances of CH$_4$ and NH$_3$ and lower abundances for CO and N$_{2}$. However, numerous BD atmospheric observations dating back to Gliese 229B have detected above equilibrium values of CO \citep{Noll1997, Oppenheimer1998} and a dearth of NH$_3$ \citep{Saumon2000,Saumon2006}. 

\par This disequilibrium was expected by atmospheric modelers based on observations of Jupiter and Saturn \citep{Fegley&Lodders1996}. Chemical disequilibrium might indicate atmospheric circulation with convection or advection transporting CO into the higher stratosphere while sequestering NH$_3$ within the BD and gas giant interiors \citep{Noll1997,Griffith&Yelle1999,Saumon2000,Leggett2007b}.

\par Applying Doppler imaging to CO, CH$_4$, and NH$_3$ spectral lines could potentially create BD and gas giant exoplanet surface maps detailing regions of increased convective circulation as well as other types of vertical transport observed in solar system stratospheres. Connecting these maps to dynamical models would allow for greater understanding of ultracool objects' atmospheric structure and dynamics. 

\section{Summary}

\par We have developed an analytical framework for inferring surface inhomogeneities on ultracool objects. The analytical framework is internally validated versus a numerical model for different inclinations and varying latitudinal and longitudinal displacements in both spectroscopic and photometric modes. This model which employs nested sampling demonstrates a computationally inexpensive and easy-to-implement approach to inferring surface inhomogeneities and is capable of retrieving spot location, size, and contrast with relatively high resolution. We then applied the analytical technique to archival spectroscopic and photometric data and retrieved similar results to that found in the literature. Moving forward, this technique can be useful for Doppler imaging the surfaces of increasingly faint ultracool objects and those with spectra made available by upcoming ELTs. Additionally, this technique could be paired with GCMs and theoretical models to discover and validate the complex dynamics driving atmospheric circulation in ultracool objects.

\section{Acknowledgements}

\par The authors would like to thank the United States Air Force Academy, Department of Physics for sponsoring the graduate work of the first author. Special thanks to the anonymous reviewer for providing insightful and constructive feedback. We would also like to thank Dr. Rodrigo Luger of the Flatiron Institute Center for Computational Astrophysics for advice on modeling stellar surfaces. Additionally, we would like to thank Dr. Theodora Karalidi for sharing the \cite{Buenzli15a} photometric data used in this article. Finally, we would like to thank the Group for Studies of Exoplanets (GFORSE) at The Ohio State University, Department of Astronomy for continuous feedback throughout the development of this research.

\par The views expressed in this article are those of the author and do not necessarily reflect the official policy or position of the Air Force, the Department of Defense, or the U.S. Government.

\software{Dynesty \citep{speagle20}, scipy \citep{scipy2020}}

\clearpage
\bibliography{references}{}

\begin{thebibliography}{}
\expandafter\ifx\csname natexlab\endcsname\relax\def\natexlab#1{#1}\fi
\providecommand{\url}[1]{\href{#1}{#1}}
\providecommand{\dodoi}[1]{doi:~\href{http://doi.org/#1}{\nolinkurl{#1}}}
\providecommand{\doeprint}[1]{\href{http://ascl.net/#1}{\nolinkurl{http://ascl.net/#1}}}
\providecommand{\doarXiv}[1]{\href{https://arxiv.org/abs/#1}{\nolinkurl{https://arxiv.org/abs/#1}}}

\bibitem[{{Ackerman} \& {Marley}(2001)}]{ackerman&marley01}
{Ackerman}, A.~S., \& {Marley}, M.~S. 2001, \apj, 556, 872,
  \dodoi{10.1086/321540}

\bibitem[{{Albrecht} {et~al.}(2007){Albrecht}, {Reffert}, {Quirrenbach},
  {Mitchell}, \& {Snellen}}]{Albrecht2007}
{Albrecht}, S., {Reffert}, S., {Quirrenbach}, A., {Mitchell}, D.~S., \&
  {Snellen}, I. 2007, in Astronomical Society of the Pacific Conference Series,
  Vol. 370, Solar and Stellar Physics Through Eclipses, ed. O.~{Demircan},
  S.~O. {Selam}, \& B.~{Albayrak}, 218

\bibitem[{{Allard} {et~al.}(2001){Allard}, {Hauschildt}, {Alexander},
  {Tamanai}, \& {Schweitzer}}]{Allard2001}
{Allard}, F., {Hauschildt}, P.~H., {Alexander}, D.~R., {Tamanai}, A., \&
  {Schweitzer}, A. 2001, \apj, 556, 357, \dodoi{10.1086/321547}

\bibitem[{{Ammons} \& {Garcia}(2019)}]{ammons&garcia19}
{Ammons}, S.~M., \& {Garcia}, V. 2019, 233, 114.07

\bibitem[{{Apai} {et~al.}(2021){Apai}, {Nardiello}, \& {Bedin}}]{apai21}
{Apai}, D., {Nardiello}, D., \& {Bedin}, L.~R. 2021, \apj, 906, 64,
  \dodoi{10.3847/1538-4357/abcb97}

\bibitem[{{Apai} {et~al.}(2013){Apai}, {Radigan}, {Buenzli}, {Burrows}, {Reid},
  \& {Jayawardhana}}]{Apai2013}
{Apai}, D., {Radigan}, J., {Buenzli}, E., {et~al.} 2013, \apj, 768, 121,
  \dodoi{10.1088/0004-637X/768/2/121}

\bibitem[{{Barman} {et~al.}(2015){Barman}, {Konopacky}, {Macintosh}, \&
  {Marois}}]{Barman2015}
{Barman}, T.~S., {Konopacky}, Q.~M., {Macintosh}, B., \& {Marois}, C. 2015,
  \apj, 804, 61, \dodoi{10.1088/0004-637X/804/1/61}

\bibitem[{{Barman} {et~al.}(2011){Barman}, {Macintosh}, {Konopacky}, \&
  {Marois}}]{Barman2011}
{Barman}, T.~S., {Macintosh}, B., {Konopacky}, Q.~M., \& {Marois}, C. 2011,
  \apj, 733, 65, \dodoi{10.1088/0004-637X/733/1/65}

\bibitem[{{Barnes} \& {Collier Cameron}(2001)}]{Barnes&CollierCameron2001b}
{Barnes}, J.~R., \& {Collier Cameron}, A. 2001, \mnras, 326, 950,
  \dodoi{10.1046/j.1365-8711.2001.04649.x}

\bibitem[{{Barnes} {et~al.}(2001){Barnes}, {Collier Cameron}, {James}, \&
  {Donati}}]{Barnes&CollierCameron2001a}
{Barnes}, J.~R., {Collier Cameron}, A., {James}, D.~J., \& {Donati}, J.~F.
  2001, \mnras, 324, 231, \dodoi{10.1046/j.1365-8711.2001.04309.x}

\bibitem[{{Barnes} {et~al.}(2004){Barnes}, {James}, \& {Collier
  Cameron}}]{Barnes2004}
{Barnes}, J.~R., {James}, D.~J., \& {Collier Cameron}, A. 2004, \mnras, 352,
  589, \dodoi{10.1111/j.1365-2966.2004.07949.x}

\bibitem[{{Barnes} {et~al.}(2017){Barnes}, {Jeffers}, {Haswell}, {Jones},
  {Shulyak}, {Pavlenko}, \& {Jenkins}}]{barnes17}
{Barnes}, J.~R., {Jeffers}, S.~V., {Haswell}, C.~A., {et~al.} 2017, \mnras,
  471, 811, \dodoi{10.1093/mnras/stx1482}

\bibitem[{{Barnes} {et~al.}(2015){Barnes}, {Jeffers}, {Jones}, {Pavlenko},
  {Jenkins}, {Haswell}, \& {Lohr}}]{barnes15}
{Barnes}, J.~R., {Jeffers}, S.~V., {Jones}, H.~R.~A., {et~al.} 2015, \apj, 812,
  42, \dodoi{10.1088/0004-637X/812/1/42}

\bibitem[{{Bedin} {et~al.}(2017){Bedin}, {Pourbaix}, {Apai}, {Burgasser},
  {Buenzli}, {Boffin}, \& {Libralato}}]{bedin17}
{Bedin}, L.~R., {Pourbaix}, D., {Apai}, D., {et~al.} 2017, \mnras, 470, 1140,
  \dodoi{10.1093/mnras/stx1177}

\bibitem[{{Biller} {et~al.}(2013){Biller}, {Crossfield}, {Mancini}, {Ciceri},
  {Southworth}, {Kopytova}, {Bonnefoy}, {Deacon}, {Schlieder}, {Buenzli},
  {Brandner}, {Allard}, {Homeier}, {Freytag}, {Bailer-Jones}, {Greiner},
  {Henning}, \& {Goldman}}]{biller13}
{Biller}, B.~A., {Crossfield}, I. J.~M., {Mancini}, L., {et~al.} 2013, \apjl,
  778, L10, \dodoi{10.1088/2041-8205/778/1/L10}

\bibitem[{{Biller} {et~al.}(2018){Biller}, {Vos}, {Buenzli}, {Allers},
  {Bonnefoy}, {Charnay}, {B{\'e}zard}, {Allard}, {Homeier}, {Bonavita},
  {Brandner}, {Crossfield}, {Dupuy}, {Henning}, {Kopytova}, {Liu},
  {Manjavacas}, \& {Schlieder}}]{Biller2018}
{Biller}, B.~A., {Vos}, J., {Buenzli}, E., {et~al.} 2018, \aj, 155, 95,
  \dodoi{10.3847/1538-3881/aaa5a6}

\bibitem[{{Buenzli} {et~al.}(2014){Buenzli}, {Apai}, {Radigan}, {Reid}, \&
  {Flateau}}]{Buenzli2014}
{Buenzli}, E., {Apai}, D., {Radigan}, J., {Reid}, I.~N., \& {Flateau}, D. 2014,
  \apj, 782, 77, \dodoi{10.1088/0004-637X/782/2/77}

\bibitem[{{Buenzli} {et~al.}(2015{\natexlab{a}}){Buenzli}, {Marley}, {Apai},
  {Saumon}, {Biller}, {Crossfield}, \& {Radigan}}]{Buenzli15b}
{Buenzli}, E., {Marley}, M.~S., {Apai}, D., {et~al.} 2015{\natexlab{a}}, \apj,
  812, 163, \dodoi{10.1088/0004-637X/812/2/163}

\bibitem[{{Buenzli} {et~al.}(2015{\natexlab{b}}){Buenzli}, {Saumon}, {Marley},
  {Apai}, {Radigan}, {Bedin}, {Reid}, \& {Morley}}]{Buenzli15a}
{Buenzli}, E., {Saumon}, D., {Marley}, M.~S., {et~al.} 2015{\natexlab{b}},
  \apj, 798, 127, \dodoi{10.1088/0004-637X/798/2/127}

\bibitem[{{Buenzli} {et~al.}(2012){Buenzli}, {Apai}, {Morley}, {Flateau},
  {Showman}, {Burrows}, {Marley}, {Lewis}, \& {Reid}}]{Buenzli2012}
{Buenzli}, E., {Apai}, D., {Morley}, C.~V., {et~al.} 2012, \apjl, 760, L31,
  \dodoi{10.1088/2041-8205/760/2/L31}

\bibitem[{{Burgasser} {et~al.}(2002){Burgasser}, {Marley}, {Ackerman},
  {Saumon}, {Lodders}, {Dahn}, {Harris}, \& {Kirkpatrick}}]{Burgasser2002}
{Burgasser}, A.~J., {Marley}, M.~S., {Ackerman}, A.~S., {et~al.} 2002, \apjl,
  571, L151, \dodoi{10.1086/341343}

\bibitem[{{Burgasser} {et~al.}(2013){Burgasser}, {Sheppard}, \&
  {Luhman}}]{burgasser13}
{Burgasser}, A.~J., {Sheppard}, S.~S., \& {Luhman}, K.~L. 2013, \apj, 772, 129,
  \dodoi{10.1088/0004-637X/772/2/129}

\bibitem[{{Burgasser} {et~al.}(2014){Burgasser}, {Gillon}, {Faherty},
  {Radigan}, {Triaud}, {Plavchan}, {Street}, {Jehin}, {Delrez}, \&
  {Opitom}}]{burgasser14}
{Burgasser}, A.~J., {Gillon}, M., {Faherty}, J.~K., {et~al.} 2014, \apj, 785,
  48, \dodoi{10.1088/0004-637X/785/1/48}

\bibitem[{{Burrows} {et~al.}(2006){Burrows}, {Sudarsky}, \&
  {Hubeny}}]{Burrows2006}
{Burrows}, A., {Sudarsky}, D., \& {Hubeny}, I. 2006, \apj, 640, 1063,
  \dodoi{10.1086/500293}

\bibitem[{{Chabrier} {et~al.}(2000){Chabrier}, {Baraffe}, {Allard}, \&
  {Hauschildt}}]{Chabrier2000}
{Chabrier}, G., {Baraffe}, I., {Allard}, F., \& {Hauschildt}, P. 2000, \apj,
  542, 464, \dodoi{10.1086/309513}

\bibitem[{{Claret} {et~al.}(2012){Claret}, {Hauschildt}, \&
  {Witte}}]{Claret2012}
{Claret}, A., {Hauschildt}, P.~H., \& {Witte}, S. 2012, \aap, 546, A14,
  \dodoi{10.1051/0004-6361/201219849}

\bibitem[{{Collier Cameron} \& {Unruh}(1994)}]{CollierCameron&Unruh1994}
{Collier Cameron}, A., \& {Unruh}, Y.~C. 1994, \mnras, 269, 814,
  \dodoi{10.1093/mnras/269.3.814}

\bibitem[{{Crossfield} {et~al.}(2014){Crossfield}, {Biller}, {Schlieder},
  {Deacon}, {Bonnefoy}, {Homeier}, {Allard}, {Buenzli}, {Henning}, {Brandner},
  {Goldman}, \& {Kopytova}}]{crossfield14}
{Crossfield}, I.~J.~M., {Biller}, B., {Schlieder}, J.~E., {et~al.} 2014, \nat,
  505, 654, \dodoi{10.1038/nature12955}

\bibitem[{{Currie} {et~al.}(2011){Currie}, {Burrows}, {Itoh}, {Matsumura},
  {Fukagawa}, {Apai}, {Madhusudhan}, {Hinz}, {Rodigas}, {Kasper}, {Pyo}, \&
  {Ogino}}]{Currie2011}
{Currie}, T., {Burrows}, A., {Itoh}, Y., {et~al.} 2011, \apj, 729, 128,
  \dodoi{10.1088/0004-637X/729/2/128}

\bibitem[{{Currie} {et~al.}(2014){Currie}, {Burrows}, {Girard}, {Cloutier},
  {Fukagawa}, {Sorahana}, {Kuchner}, {Kenyon}, {Madhusudhan}, {Itoh},
  {Jayawardhana}, {Matsumura}, \& {Pyo}}]{Currie2014}
{Currie}, T., {Burrows}, A., {Girard}, J.~H., {et~al.} 2014, \apj, 795, 133,
  \dodoi{10.1088/0004-637X/795/2/133}

\bibitem[{{Cushing} {et~al.}(2008){Cushing}, {Marley}, {Saumon}, {Kelly},
  {Vacca}, {Rayner}, {Freedman}, {Lodders}, \& {Roellig}}]{Cushing2008}
{Cushing}, M.~C., {Marley}, M.~S., {Saumon}, D., {et~al.} 2008, \apj, 678,
  1372, \dodoi{10.1086/526489}

\bibitem[{{Deutsch}(1958)}]{Deutsch1958}
{Deutsch}, A.~J. 1958, in Electromagnetic Phenomena in Cosmical Physics, ed.
  B.~{Lehnert}, Vol.~6, 209

\bibitem[{{Deutsch}(1970)}]{Deutsch1970}
{Deutsch}, A.~J. 1970, \apj, 159, 985, \dodoi{10.1086/150376}

\bibitem[{Donatelli \& Reichel(2014)}]{donatelli&reichel14}
Donatelli, M., \& Reichel, L. 2014, Journal of Computational and Applied
  Mathematics, 272, 334

\bibitem[{{Donati}(1999)}]{Donati1999}
{Donati}, J.~F. 1999, \mnras, 302, 457,
  \dodoi{10.1046/j.1365-8711.1999.02096.x}

\bibitem[{{Donati} {et~al.}(1997){Donati}, {Semel}, {Carter}, {Rees}, \&
  {Collier Cameron}}]{Donati1997}
{Donati}, J.~F., {Semel}, M., {Carter}, B.~D., {Rees}, D.~E., \& {Collier
  Cameron}, A. 1997, \mnras, 291, 658, \dodoi{10.1093/mnras/291.4.658}

\bibitem[{{Dupuy} \& {Liu}(2017)}]{Dupuy2017}
{Dupuy}, T.~J., \& {Liu}, M.~C. 2017, \apjs, 231, 15,
  \dodoi{10.3847/1538-4365/aa5e4c}

\bibitem[{{Enoch} {et~al.}(2003){Enoch}, {Brown}, \& {Burgasser}}]{Enoch2003}
{Enoch}, M.~L., {Brown}, M.~E., \& {Burgasser}, A.~J. 2003, \aj, 126, 1006,
  \dodoi{10.1086/376598}

\bibitem[{{Eriksson} {et~al.}(2019){Eriksson}, {Janson}, \&
  {Calissendorff}}]{Eriksson2019}
{Eriksson}, S.~C., {Janson}, M., \& {Calissendorff}, P. 2019, \aap, 629, A145,
  \dodoi{10.1051/0004-6361/201935671}

\bibitem[{{Falk} \& {Wehlau}(1974)}]{Falk1974}
{Falk}, A.~E., \& {Wehlau}, W.~H. 1974, \apj, 192, 409, \dodoi{10.1086/153072}

\bibitem[{{Fegley} \& {Lodders}(1996)}]{Fegley&Lodders1996}
{Fegley}, Bruce, J., \& {Lodders}, K. 1996, \apjl, 472, L37,
  \dodoi{10.1086/310356}

\bibitem[{{Finociety} {et~al.}(2021){Finociety}, {Donati}, {Klein}, {Zaire},
  {Lehmann}, {Moutou}, {Bouvier}, {Alencar}, {Yu}, {Grankin}, {Artigau},
  {Doyon}, {Delfosse}, {Fouqu{\'e}}, {H{\'e}brard}, {Jardine},
  {K{\'o}sp{\'a}l}, {M{\'e}nard}, {M{\'e}nard}, \& {SLS
  consortium}}]{Finociety2021}
{Finociety}, B., {Donati}, J.~F., {Klein}, B., {et~al.} 2021, \mnras, 508,
  3427, \dodoi{10.1093/mnras/stab2778}

\bibitem[{{Freytag} {et~al.}(2010){Freytag}, {Allard}, {Ludwig}, {Homeier}, \&
  {Steffen}}]{Freytag2010}
{Freytag}, B., {Allard}, F., {Ludwig}, H.~G., {Homeier}, D., \& {Steffen}, M.
  2010, \aap, 513, A19, \dodoi{10.1051/0004-6361/200913354}

\bibitem[{{Garcia} {et~al.}(2017){Garcia}, {Ammons}, {Salama}, {Crossfield},
  {Bendek}, {Chilcote}, {Garrel}, {Graham}, {Kalas}, {Konopacky}, {Lu},
  {Macintosh}, {Marin}, {Marois}, {Nielsen}, {Neichel}, {Pham}, {De Rosa},
  {Ryan}, {Service}, \& {Sivo}}]{garcia17}
{Garcia}, E.~V., {Ammons}, S.~M., {Salama}, M., {et~al.} 2017, \apj, 846, 97,
  \dodoi{10.3847/1538-4357/aa844f}

\bibitem[{{Gillon} {et~al.}(2013){Gillon}, {Triaud}, {Jehin}, {Delrez},
  {Opitom}, {Magain}, {Lendl}, \& {Queloz}}]{gillon13}
{Gillon}, M., {Triaud}, A.~H.~M.~J., {Jehin}, E., {et~al.} 2013, \aap, 555, L5,
  \dodoi{10.1051/0004-6361/201321620}

\bibitem[{{Goncharskii} {et~al.}(1982){Goncharskii}, {Stepanov}, {Khokhlova},
  \& {Yagola}}]{Goncharskii1982}
{Goncharskii}, A.~V., {Stepanov}, V.~V., {Khokhlova}, V.~L., \& {Yagola}, A.~G.
  1982, \sovast, 26, 690

\bibitem[{{Goncharskii} {et~al.}(1977){Goncharskii}, {Stepanov}, {Kokhlova}, \&
  {Yagola}}]{Goncharskii1977}
{Goncharskii}, A.~V., {Stepanov}, V.~V., {Kokhlova}, V.~L., \& {Yagola}, A.~G.
  1977, Pisma v Astronomicheskii Zhurnal, 3, 278

\bibitem[{{Gray}(2008)}]{gray22}
{Gray}, D.~F. 2008

\bibitem[{{Griffith} \& {Yelle}(1999)}]{Griffith&Yelle1999}
{Griffith}, C.~A., \& {Yelle}, R.~V. 1999, \apjl, 519, L85,
  \dodoi{10.1086/312103}

\bibitem[{{G{\"u}nther} {et~al.}(2020){G{\"u}nther}, {Zhan}, {Seager},
  {Rimmer}, {Ranjan}, {Stassun}, {Oelkers}, {Daylan}, {Newton}, {Kristiansen},
  {Olah}, {Gillen}, {Rappaport}, {Ricker}, {Vanderspek}, {Latham}, {Winn},
  {Jenkins}, {Glidden}, {Fausnaugh}, {Levine}, {Dittmann}, {Quinn},
  {Krishnamurthy}, \& {Ting}}]{gunther20}
{G{\"u}nther}, M.~N., {Zhan}, Z., {Seager}, S., {et~al.} 2020, \aj, 159, 60,
  \dodoi{10.3847/1538-3881/ab5d3a}

\bibitem[{{Hatzes}(1995)}]{Hatzes1995}
{Hatzes}, A.~P. 1995, \apj, 451, 784, \dodoi{10.1086/176265}

\bibitem[{{Heinze} {et~al.}(2021){Heinze}, {Metchev}, {Kurtev}, \&
  {Gillon}}]{heinze21}
{Heinze}, A.~N., {Metchev}, S., {Kurtev}, R., \& {Gillon}, M. 2021, \apj, 920,
  108, \dodoi{10.3847/1538-4357/ac178b}

\bibitem[{{Hirano} {et~al.}(2011){Hirano}, {Suto}, {Winn}, {Taruya}, {Narita},
  {Albrecht}, \& {Sato}}]{Hirano2011}
{Hirano}, T., {Suto}, Y., {Winn}, J.~N., {et~al.} 2011, \apj, 742, 69,
  \dodoi{10.1088/0004-637X/742/2/69}

\bibitem[{Ilin {et~al.}(2021)Ilin, Poppenhaeger, Schmidt, Jarvinen, Newton,
  Alvarado-Gomez, Pineda, Davenport, Oshagh, \& Ilyin}]{ilin21}
Ilin, E., Poppenhaeger, K., Schmidt, S.~J., {et~al.} 2021, Monthly Notices of
  the Royal Astronomical Society, Accepted

\bibitem[{{J{\"a}rvinen} {et~al.}(2008){J{\"a}rvinen}, {Korhonen},
  {Berdyugina}, {Ilyin}, {Strassmeier}, {Weber}, {Savanov}, \&
  {Tuominen}}]{jarvinen08}
{J{\"a}rvinen}, S.~P., {Korhonen}, H., {Berdyugina}, S.~V., {et~al.} 2008,
  \aap, 488, 1047, \dodoi{10.1051/0004-6361:200809837}

\bibitem[{{Johnson} {et~al.}(2014){Johnson}, {Cochran}, {Albrecht},
  {Dodson-Robinson}, {Winn}, \& {Gullikson}}]{Johnson2014}
{Johnson}, M.~C., {Cochran}, W.~D., {Albrecht}, S., {et~al.} 2014, \apj, 790,
  30, \dodoi{10.1088/0004-637X/790/1/30}

\bibitem[{{Jones} \& {Tsuji}(1997)}]{Jones&Tsuji1997}
{Jones}, H. R.~A., \& {Tsuji}, T. 1997, \apjl, 480, L39, \dodoi{10.1086/310619}

\bibitem[{{Karalidi} {et~al.}(2016){Karalidi}, {Apai}, {Marley}, \&
  {Buenzli}}]{karalidi16}
{Karalidi}, T., {Apai}, D., {Marley}, M.~S., \& {Buenzli}, E. 2016, \apj, 825,
  90, \dodoi{10.3847/0004-637X/825/2/90}

\bibitem[{{Karalidi} {et~al.}(2015){Karalidi}, {Apai}, {Schneider}, {Hanson},
  \& {Pasachoff}}]{karalidi15}
{Karalidi}, T., {Apai}, D., {Schneider}, G., {Hanson}, J.~R., \& {Pasachoff},
  J.~M. 2015, \apj, 814, 65, \dodoi{10.1088/0004-637X/814/1/65}

\bibitem[{Kass \& Raftery(1995)}]{Kass&Raftery1995}
Kass, R.~E., \& Raftery, A.~E. 1995, Journal of the American Statistical
  Association, 90, 773, \dodoi{10.1080/01621459.1995.10476572}

\bibitem[{{Kasting} {et~al.}(1993){Kasting}, {Whitmire}, \&
  {Reynolds}}]{kasting93}
{Kasting}, J.~F., {Whitmire}, D.~P., \& {Reynolds}, R.~T. 1993, \icarus, 101,
  108, \dodoi{10.1006/icar.1993.1010}

\bibitem[{{Khokhlova}(1976)}]{Khokhlova1976}
{Khokhlova}, V.~L. 1976, \sovast, 19, 576

\bibitem[{{Kniazev} {et~al.}(2013){Kniazev}, {Vaisanen}, {Mu{\v{z}}i{\'c}},
  {Mehner}, {Boffin}, {Kurtev}, {Melo}, {Ivanov}, {Girard}, {Mawet},
  {Schmidtobreick}, {Huelamo}, {Borissova}, {Minniti}, {Ishibashi}, {Potter},
  {Beletsky}, {Buckley}, {Crawford}, {Gulbis}, {Kotze}, {Miszalski},
  {Pickering}, {Romero Colmenero}, \& {Williams}}]{kniazev13}
{Kniazev}, A.~Y., {Vaisanen}, P., {Mu{\v{z}}i{\'c}}, K., {et~al.} 2013, \apj,
  770, 124, \dodoi{10.1088/0004-637X/770/2/124}

\bibitem[{{Kochukhov} {et~al.}(2010){Kochukhov}, {Makaganiuk}, \&
  {Piskunov}}]{kochokhov10}
{Kochukhov}, O., {Makaganiuk}, V., \& {Piskunov}, N. 2010, \aap, 524, A5,
  \dodoi{10.1051/0004-6361/201015429}

\bibitem[{{Konopacky} {et~al.}(2013){Konopacky}, {Barman}, {Macintosh}, \&
  {Marois}}]{Konopacky2013}
{Konopacky}, Q.~M., {Barman}, T.~S., {Macintosh}, B.~A., \& {Marois}, C. 2013,
  Science, 339, 1398, \dodoi{10.1126/science.1232003}

\bibitem[{{Kopparapu} {et~al.}(2013){Kopparapu}, {Ramirez}, {Kasting}, {Eymet},
  {Robinson}, {Mahadevan}, {Terrien}, {Domagal-Goldman}, {Meadows}, \&
  {Deshpande}}]{kopparapu13}
{Kopparapu}, R.~K., {Ramirez}, R., {Kasting}, J.~F., {et~al.} 2013, \apj, 765,
  131, \dodoi{10.1088/0004-637X/765/2/131}

\bibitem[{{Lavie} {et~al.}(2017){Lavie}, {Mendon{\c{c}}a}, {Mordasini},
  {Malik}, {Bonnefoy}, {Demory}, {Oreshenko}, {Grimm}, {Ehrenreich}, \&
  {Heng}}]{Lavie2017}
{Lavie}, B., {Mendon{\c{c}}a}, J.~M., {Mordasini}, C., {et~al.} 2017, \aj, 154,
  91, \dodoi{10.3847/1538-3881/aa7ed8}

\bibitem[{{Lef{\`e}vre} {et~al.}(2022){Lef{\`e}vre}, {Tan}, {Lee}, \&
  {Pierrehumbert}}]{Lefevre2022}
{Lef{\`e}vre}, M., {Tan}, X., {Lee}, E. K.~H., \& {Pierrehumbert}, R.~T. 2022,
  arXiv e-prints, arXiv:2203.08625.
\newblock \doarXiv{2203.08625}

\bibitem[{{Leggett} {et~al.}(2007){Leggett}, {Marley}, {Freedman}, {Saumon},
  {Liu}, {Geballe}, {Golimowski}, \& {Stephens}}]{Leggett2007b}
{Leggett}, S.~K., {Marley}, M.~S., {Freedman}, R., {et~al.} 2007, \apj, 667,
  537, \dodoi{10.1086/519948}

\bibitem[{{Lew} {et~al.}(2016){Lew}, {Apai}, {Zhou}, {Schneider}, {Burgasser},
  {Karalidi}, {Yang}, {Marley}, {Cowan}, {Bedin}, {Metchev}, {Radigan}, \&
  {Lowrance}}]{Lew2016}
{Lew}, B. W.~P., {Apai}, D., {Zhou}, Y., {et~al.} 2016, \apjl, 829, L32,
  \dodoi{10.3847/2041-8205/829/2/L32}

\bibitem[{{Lew} {et~al.}(2020{\natexlab{a}}){Lew}, {Apai}, {Zhou}, {Radigan},
  {Marley}, {Schneider}, {Cowan}, {Miles-P{\'a}ez}, {Manjavacas}, {Karalidi},
  {Bedin}, {Lowrance}, \& {Burgasser}}]{Lew2020a}
---. 2020{\natexlab{a}}, \aj, 159, 125, \dodoi{10.3847/1538-3881/ab5f59}

\bibitem[{{Lew} {et~al.}(2020{\natexlab{b}}){Lew}, {Apai}, {Marley}, {Saumon},
  {Schneider}, {Zhou}, {Cowan}, {Karalidi}, {Manjavacas}, {Bedin}, \&
  {Miles-P{\'a}ez}}]{Lew2020b}
{Lew}, B. W.~P., {Apai}, D., {Marley}, M., {et~al.} 2020{\natexlab{b}}, \apj,
  903, 15, \dodoi{10.3847/1538-4357/abb81d}

\bibitem[{{Luger} {et~al.}(2019){Luger}, {Agol}, {Foreman-Mackey}, {Fleming},
  {Lustig-Yaeger}, \& {Deitrick}}]{Luger2019}
{Luger}, R., {Agol}, E., {Foreman-Mackey}, D., {et~al.} 2019, \aj, 157, 64,
  \dodoi{10.3847/1538-3881/aae8e5}

\bibitem[{{Luger} {et~al.}(2021){Luger}, {Bedell}, {Foreman-Mackey},
  {Crossfield}, {Zhao}, \& {Hogg}}]{luger21a}
{Luger}, R., {Bedell}, M., {Foreman-Mackey}, D., {et~al.} 2021, arXiv e-prints,
  arXiv:2110.06271.
\newblock \doarXiv{2110.06271}

\bibitem[{{Luhman}(2013)}]{luhman13}
{Luhman}, K.~L. 2013, \apjl, 767, L1, \dodoi{10.1088/2041-8205/767/1/L1}

\bibitem[{{Madhusudhan} {et~al.}(2011){Madhusudhan}, {Burrows}, \&
  {Currie}}]{Madhusudhan2011}
{Madhusudhan}, N., {Burrows}, A., \& {Currie}, T. 2011, \apj, 737, 34,
  \dodoi{10.1088/0004-637X/737/1/34}

\bibitem[{{Manjavacas} {et~al.}(2021){Manjavacas}, {Karalidi}, {Vos}, {Biller},
  \& {Lew}}]{Manjavacas2021}
{Manjavacas}, E., {Karalidi}, T., {Vos}, J.~M., {Biller}, B.~A., \& {Lew}, B.
  W.~P. 2021, \aj, 162, 179, \dodoi{10.3847/1538-3881/ac174c}

\bibitem[{{Manjavacas} {et~al.}(2018){Manjavacas}, {Apai}, {Zhou}, {Karalidi},
  {Lew}, {Schneider}, {Cowan}, {Metchev}, {Miles-P{\'a}ez}, {Burgasser},
  {Radigan}, {Bedin}, {Lowrance}, \& {Marley}}]{Manjavacas2018}
{Manjavacas}, E., {Apai}, D., {Zhou}, Y., {et~al.} 2018, \aj, 155, 11,
  \dodoi{10.3847/1538-3881/aa984f}

\bibitem[{{Manjavacas} {et~al.}(2019{\natexlab{a}}){Manjavacas}, {Apai},
  {Zhou}, {Lew}, {Schneider}, {Metchev}, {Miles-P{\'a}ez}, {Radigan}, {Marley},
  {Cowan}, {Karalidi}, {Burgasser}, {Bedin}, {Lowrance}, \&
  {Kauffmann}}]{Manjavacas2019a}
---. 2019{\natexlab{a}}, \aj, 157, 101, \dodoi{10.3847/1538-3881/aaf88f}

\bibitem[{{Manjavacas} {et~al.}(2019{\natexlab{b}}){Manjavacas}, {Apai}, {Lew},
  {Zhou}, {Schneider}, {Burgasser}, {Karalidi}, {Miles-P{\'a}ez}, {Lowrance},
  {Cowan}, {Bedin}, {Marley}, {Metchev}, \& {Radigan}}]{Manjavacas2019b}
{Manjavacas}, E., {Apai}, D., {Lew}, B. W.~P., {et~al.} 2019{\natexlab{b}},
  \apjl, 875, L15, \dodoi{10.3847/2041-8213/ab13b9}

\bibitem[{{Marley} {et~al.}(2012){Marley}, {Saumon}, {Cushing}, {Ackerman},
  {Fortney}, \& {Freedman}}]{Marley2012}
{Marley}, M.~S., {Saumon}, D., {Cushing}, M., {et~al.} 2012, \apj, 754, 135,
  \dodoi{10.1088/0004-637X/754/2/135}

\bibitem[{{Marley} {et~al.}(2010){Marley}, {Saumon}, \&
  {Goldblatt}}]{Marley2010}
{Marley}, M.~S., {Saumon}, D., \& {Goldblatt}, C. 2010, \apjl, 723, L117,
  \dodoi{10.1088/2041-8205/723/1/L117}

\bibitem[{{Marley} {et~al.}(2002){Marley}, {Seager}, {Saumon}, {Lodders},
  {Ackerman}, {Freedman}, \& {Fan}}]{Marley2002}
{Marley}, M.~S., {Seager}, S., {Saumon}, D., {et~al.} 2002, \apj, 568, 335,
  \dodoi{10.1086/338800}

\bibitem[{{Metchev} {et~al.}(2015){Metchev}, {Heinze}, {Apai}, {Flateau},
  {Radigan}, {Burgasser}, {Marley}, {Artigau}, {Plavchan}, \&
  {Goldman}}]{Metchev2015}
{Metchev}, S.~A., {Heinze}, A., {Apai}, D., {et~al.} 2015, \apj, 799, 154,
  \dodoi{10.1088/0004-637X/799/2/154}

\bibitem[{{Miles-P{\'a}ez} {et~al.}(2019){Miles-P{\'a}ez}, {Metchev}, {Apai},
  {Zhou}, {Manjavacas}, {Karalidi}, {Lew}, {Burgasser}, {Bedin}, {Cowan},
  {Lowrance}, {Marley}, {Radigan}, \& {Schneider}}]{MilesPaez2019}
{Miles-P{\'a}ez}, P.~A., {Metchev}, S., {Apai}, D., {et~al.} 2019, \apj, 883,
  181, \dodoi{10.3847/1538-4357/ab3d25}

\bibitem[{{Molli{\`e}re} {et~al.}(2020){Molli{\`e}re}, {Stolker}, {Lacour},
  {Otten}, {Shangguan}, {Charnay}, {Molyarova}, {Nowak}, {Henning}, {Marleau},
  {Semenov}, {van Dishoeck}, {Eisenhauer}, {Garcia}, {Garcia Lopez}, {Girard},
  {Greenbaum}, {Hinkley}, {Kervella}, {Kreidberg}, {Maire}, {Nasedkin},
  {Pueyo}, {Snellen}, {Vigan}, {Wang}, {de Zeeuw}, \& {Zurlo}}]{Molliere2020}
{Molli{\`e}re}, P., {Stolker}, T., {Lacour}, S., {et~al.} 2020, \aap, 640,
  A131, \dodoi{10.1051/0004-6361/202038325}

\bibitem[{{Mulders} {et~al.}(2015){Mulders}, {Pascucci}, \&
  {Apai}}]{Mulders2015b}
{Mulders}, G.~D., {Pascucci}, I., \& {Apai}, D. 2015, \apj, 814, 130,
  \dodoi{10.1088/0004-637X/814/2/130}

\bibitem[{{Noll} {et~al.}(2000){Noll}, {Geballe}, {Leggett}, \&
  {Marley}}]{Noll2000}
{Noll}, K.~S., {Geballe}, T.~R., {Leggett}, S.~K., \& {Marley}, M.~S. 2000,
  \apjl, 541, L75, \dodoi{10.1086/312906}

\bibitem[{{Noll} {et~al.}(1997){Noll}, {Geballe}, \& {Marley}}]{Noll1997}
{Noll}, K.~S., {Geballe}, T.~R., \& {Marley}, M.~S. 1997, \apjl, 489, L87,
  \dodoi{10.1086/310954}

\bibitem[{{Oppenheimer} {et~al.}(1998){Oppenheimer}, {Kulkarni}, {Matthews}, \&
  {van Kerkwijk}}]{Oppenheimer1998}
{Oppenheimer}, B.~R., {Kulkarni}, S.~R., {Matthews}, K., \& {van Kerkwijk},
  M.~H. 1998, \apj, 502, 932, \dodoi{10.1086/305928}

\bibitem[{{Pai Asnodkar} {et~al.}(2022){Pai Asnodkar}, {Wang}, {Gaudi},
  {Cauley}, {Eastman}, {Ilyin}, {Strassmeier}, \& {Beatty}}]{PaiAsnodkar22}
{Pai Asnodkar}, A., {Wang}, J., {Gaudi}, B.~S., {et~al.} 2022, \aj, 163, 40,
  \dodoi{10.3847/1538-3881/ac32c7}

\bibitem[{{Paudel} {et~al.}(2020){Paudel}, {Gizis}, {Mullan}, {Schmidt},
  {Burgasser}, \& {Williams}}]{Paudel2020}
{Paudel}, R.~R., {Gizis}, J.~E., {Mullan}, D.~J., {et~al.} 2020, \mnras, 494,
  5751, \dodoi{10.1093/mnras/staa1137}

\bibitem[{{Paudel} {et~al.}(2018){Paudel}, {Gizis}, {Mullan}, {Schmidt},
  {Burgasser}, {Williams}, \& {Berger}}]{Paudel2018}
---. 2018, \apj, 858, 55, \dodoi{10.3847/1538-4357/aab8fe}

\bibitem[{{Piskunov} {et~al.}(1990){Piskunov}, {Tuominen}, \&
  {Vilhu}}]{Piskunov1990}
{Piskunov}, N.~E., {Tuominen}, I., \& {Vilhu}, O. 1990, \aap, 230, 363

\bibitem[{{Radigan} {et~al.}(2012){Radigan}, {Jayawardhana}, {Lafreni{\`e}re},
  {Artigau}, {Marley}, \& {Saumon}}]{Radigan2012}
{Radigan}, J., {Jayawardhana}, R., {Lafreni{\`e}re}, D., {et~al.} 2012, \apj,
  750, 105, \dodoi{10.1088/0004-637X/750/2/105}

\bibitem[{{Radigan} {et~al.}(2014){Radigan}, {Lafreni{\`e}re}, {Jayawardhana},
  \& {Artigau}}]{Radigan2014a}
{Radigan}, J., {Lafreni{\`e}re}, D., {Jayawardhana}, R., \& {Artigau}, E. 2014,
  \apj, 793, 75, \dodoi{10.1088/0004-637X/793/2/75}

\bibitem[{{Reiners} \& {Basri}(2008)}]{Reimers2008}
{Reiners}, A., \& {Basri}, G. 2008, \apj, 684, 1390, \dodoi{10.1086/590073}

\bibitem[{Rimmer {et~al.}(2018)Rimmer, Xu, Thompson, Gillen, Sutherland, \&
  Queloz}]{rimmer18}
Rimmer, P.~B., Xu, J., Thompson, S.~J., {et~al.} 2018, Science Advances, 4,
  eear3302

\bibitem[{{Sabotta} {et~al.}(2021){Sabotta}, {Schlecker}, {Chaturvedi},
  {Guenther}, {Mu{\~n}oz Rodr{\'\i}guez}, {Mu{\~n}oz S{\'a}nchez}, {Caballero},
  {Shan}, {Reffert}, {Ribas}, {Reiners}, {Hatzes}, {Amado}, {Klahr}, {Morales},
  {Quirrenbach}, {Henning}, {Dreizler}, {Pall{\'e}}, {Perger}, {Azzaro},
  {Jeffers}, {Kaminski}, {K{\"u}rster}, {Lafarga}, {Montes}, {Passegger}, \&
  {Zechmeister}}]{Sabotta2021}
{Sabotta}, S., {Schlecker}, M., {Chaturvedi}, P., {et~al.} 2021, \aap, 653,
  A114, \dodoi{10.1051/0004-6361/202140968}

\bibitem[{{Saumon} {et~al.}(2000){Saumon}, {Geballe}, {Leggett}, {Marley},
  {Freedman}, {Lodders}, {Fegley}, \& {Sengupta}}]{Saumon2000}
{Saumon}, D., {Geballe}, T.~R., {Leggett}, S.~K., {et~al.} 2000, \apj, 541,
  374, \dodoi{10.1086/309410}

\bibitem[{{Saumon} {et~al.}(2006){Saumon}, {Marley}, {Cushing}, {Leggett},
  {Roellig}, {Lodders}, \& {Freedman}}]{Saumon2006}
{Saumon}, D., {Marley}, M.~S., {Cushing}, M.~C., {et~al.} 2006, \apj, 647, 552,
  \dodoi{10.1086/505419}

\bibitem[{{Scalo} {et~al.}(2007){Scalo}, {Kaltenegger}, {Segura}, {Fridlund},
  {Ribas}, {Kulikov}, {Grenfell}, {Rauer}, {Odert}, {Leitzinger}, {Selsis},
  {Khodachenko}, {Eiroa}, {Kasting}, \& {Lammer}}]{scalo07}
{Scalo}, J., {Kaltenegger}, L., {Segura}, A.~G., {et~al.} 2007, Astrobiology,
  7, 85, \dodoi{10.1089/ast.2006.0125}

\bibitem[{{Schlawin} {et~al.}(2017){Schlawin}, {Burgasser}, {Karalidi},
  {Gizis}, \& {Teske}}]{Schlawin2017}
{Schlawin}, E., {Burgasser}, A.~J., {Karalidi}, T., {Gizis}, J.~E., \& {Teske},
  J. 2017, \apj, 849, 163, \dodoi{10.3847/1538-4357/aa90b8}

\bibitem[{{Showman} \& {Kaspi}(2013)}]{showman&kaspi13}
{Showman}, A.~P., \& {Kaspi}, Y. 2013, \apj, 776, 85,
  \dodoi{10.1088/0004-637X/776/2/85}

\bibitem[{{Showman} {et~al.}(2020){Showman}, {Tan}, \&
  {Parmentier}}]{showman20}
{Showman}, A.~P., {Tan}, X., \& {Parmentier}, V. 2020, \ssr, 216, 139,
  \dodoi{10.1007/s11214-020-00758-8}

\bibitem[{{Showman} {et~al.}(2019){Showman}, {Tan}, \& {Zhang}}]{showman19}
{Showman}, A.~P., {Tan}, X., \& {Zhang}, X. 2019, \apj, 883, 4,
  \dodoi{10.3847/1538-4357/ab384a}

\bibitem[{{Shuster}(1993)}]{Shuster1993}
{Shuster}, M.~D. 1993, IEEE Transactions on Aerospace Electronic Systems, 29,
  263, \dodoi{10.1109/7.249140}

\bibitem[{{Skemer} {et~al.}(2014){Skemer}, {Marley}, {Hinz}, {Morzinski},
  {Skrutskie}, {Leisenring}, {Close}, {Saumon}, {Bailey}, {Briguglio},
  {Defrere}, {Esposito}, {Follette}, {Hill}, {Males}, {Puglisi}, {Rodigas}, \&
  {Xompero}}]{Skemer2014}
{Skemer}, A.~J., {Marley}, M.~S., {Hinz}, P.~M., {et~al.} 2014, \apj, 792, 17,
  \dodoi{10.1088/0004-637X/792/1/17}

\bibitem[{{Skilling}(2004)}]{skilling04}
{Skilling}, J. 2004, 735, 395, \dodoi{10.1063/1.1835238}

\bibitem[{{Skilling}(2006)}]{skilling06}
---. 2006, Bayesian Analysis, 1, 833, \dodoi{10.1214/06-BA127}

\bibitem[{{Speagle}(2020)}]{speagle20}
{Speagle}, J.~S. 2020, \mnras, 493, 3132, \dodoi{10.1093/mnras/staa278}

\bibitem[{{Strassmeier}(1999)}]{Strassmeier1999b}
{Strassmeier}, K.~G. 1999, \aap, 347, 225

\bibitem[{{Strassmeier} {et~al.}(1998){Strassmeier}, {Bartus}, {Kovari},
  {Weber}, \& {Washuettl}}]{Strassmeier1998}
{Strassmeier}, K.~G., {Bartus}, J., {Kovari}, Z., {Weber}, M., \& {Washuettl},
  A. 1998, \aap, 336, 587

\bibitem[{{Strassmeier} {et~al.}(1999){Strassmeier}, {Lupinek}, {Dempsey}, \&
  {Rice}}]{Strassmeier1999a}
{Strassmeier}, K.~G., {Lupinek}, S., {Dempsey}, R.~C., \& {Rice}, J.~B. 1999,
  \aap, 347, 212

\bibitem[{{Strassmeier} \& {Rice}(1998)}]{strassmeier&rice98}
{Strassmeier}, K.~G., \& {Rice}, J.~B. 1998, \aap, 339, 497

\bibitem[{{Tan}(2022)}]{tan22}
{Tan}, X. 2022, \mnras, \dodoi{10.1093/mnras/stac344}

\bibitem[{{Tan} \& {Showman}(2019)}]{Tan&Showman19}
{Tan}, X., \& {Showman}, A.~P. 2019, \apj, 874, 111,
  \dodoi{10.3847/1538-4357/ab0c07}

\bibitem[{{Tan} \& {Showman}(2021{\natexlab{a}})}]{tan21a}
---. 2021{\natexlab{a}}, \mnras, 502, 678, \dodoi{10.1093/mnras/stab060}

\bibitem[{{Tan} \& {Showman}(2021{\natexlab{b}})}]{tan21b}
---. 2021{\natexlab{b}}, \mnras, 502, 2198, \dodoi{10.1093/mnras/stab097}

\bibitem[{{Tarter} {et~al.}(2007){Tarter}, {Backus}, {Mancinelli}, {Aurnou},
  {Backman}, {Basri}, {Boss}, {Clarke}, {Deming}, {Doyle}, {Feigelson},
  {Freund}, {Grinspoon}, {Haberle}, {Hauck}, {Heath}, {Henry}, {Hollingsworth},
  {Joshi}, {Kilston}, {Liu}, {Meikle}, {Reid}, {Rothschild}, {Scalo}, {Segura},
  {Tang}, {Tiedje}, {Turnbull}, {Walkowicz}, {Weber}, \& {Young}}]{tarter07}
{Tarter}, J.~C., {Backus}, P.~R., {Mancinelli}, R.~L., {et~al.} 2007,
  Astrobiology, 7, 30, \dodoi{10.1089/ast.2006.0124}

\bibitem[{{Tsuji} {et~al.}(1996){Tsuji}, {Ohnaka}, {Aoki}, \&
  {Nakajima}}]{Tsuji1996}
{Tsuji}, T., {Ohnaka}, K., {Aoki}, W., \& {Nakajima}, T. 1996, \aap, 308, L29

\bibitem[{{Tuomi} {et~al.}(2019){Tuomi}, {Jones}, {Butler}, {Arriagada},
  {Vogt}, {Burt}, {Laughlin}, {Holden}, {Shectman}, {Crane}, {Thompson},
  {Keiser}, {Jenkins}, {Berdi{\~n}as}, {Diaz}, {Kiraga}, \&
  {Barnes}}]{Tuomi2019}
{Tuomi}, M., {Jones}, H.~R.~A., {Butler}, R.~P., {et~al.} 2019, arXiv e-prints,
  arXiv:1906.04644.
\newblock \doarXiv{1906.04644}

\bibitem[{Virtanen {et~al.}(2020)Virtanen, Gommers, Oliphant, Haberland, Reddy,
  Cournapeau, Burovski, Peterson, Weckesser, Bright, {van der Walt}, Brett,
  Wilson, Millman, Mayorov, Nelson, Jones, Kern, Larson, Carey, Polat, Feng,
  Moore, {VanderPlas}, Laxalde, Perktold, Cimrman, Henriksen, Quintero, Harris,
  Archibald, Ribeiro, Pedregosa, {van Mulbregt}, \& {SciPy 1.0
  Contributors}}]{scipy2020}
Virtanen, P., Gommers, R., Oliphant, T.~E., {et~al.} 2020, Nature Methods, 17,
  261, \dodoi{10.1038/s41592-019-0686-2}

\bibitem[{{Vogt} \& {Penrod}(1983)}]{Vogt&Penrod1983}
{Vogt}, S.~S., \& {Penrod}, G.~D. 1983, \pasp, 95, 565, \dodoi{10.1086/131208}

\bibitem[{{Vogt} {et~al.}(1987){Vogt}, {Penrod}, \& {Hatzes}}]{vogt87}
{Vogt}, S.~S., {Penrod}, G.~D., \& {Hatzes}, A.~P. 1987, \apj, 321, 496,
  \dodoi{10.1086/165647}

\bibitem[{{Wang} {et~al.}(2018){Wang}, {David}, {Hillenbrand}, {Mawet},
  {Albrecht}, \& {Liu}}]{Wang18}
{Wang}, J., {David}, T.~J., {Hillenbrand}, L.~A., {et~al.} 2018, \apj, 865,
  141, \dodoi{10.3847/1538-4357/aadee8}

\bibitem[{{Wang} {et~al.}(2017){Wang}, {Prato}, \& {Mawet}}]{Wang17}
{Wang}, J., {Prato}, L., \& {Mawet}, D. 2017, \apj, 838, 35,
  \dodoi{10.3847/1538-4357/aa6345}

\bibitem[{{Wang} {et~al.}(2020){Wang}, {Wang}, {Ma}, {Chilcote}, {Ertel},
  {Guyon}, {Ilyin}, {Jovanovic}, {Kalas}, {Lozi}, {Macintosh}, {Strassmeier},
  \& {Stone}}]{Wang2020}
{Wang}, J., {Wang}, J.~J., {Ma}, B., {et~al.} 2020, \aj, 160, 150,
  \dodoi{10.3847/1538-3881/ababa7}

\bibitem[{{Yang} {et~al.}(2016){Yang}, {Apai}, {Marley}, {Karalidi}, {Flateau},
  {Showman}, {Metchev}, {Buenzli}, {Radigan}, {Artigau}, {Lowrance}, \&
  {Burgasser}}]{Yang2016}
{Yang}, H., {Apai}, D., {Marley}, M.~S., {et~al.} 2016, \apj, 826, 8,
  \dodoi{10.3847/0004-637X/826/1/8}

\bibitem[{{Zaire} {et~al.}(2021){Zaire}, {Donati}, \& {Klein}}]{Zaire2021}
{Zaire}, B., {Donati}, J.~F., \& {Klein}, B. 2021, \mnras, 504, 1969,
  \dodoi{10.1093/mnras/stab1019}

\bibitem[{{Zendejas} {et~al.}(2010){Zendejas}, {Segura}, \&
  {Raga}}]{zendejas10}
{Zendejas}, J., {Segura}, A., \& {Raga}, A.~C. 2010, \icarus, 210, 539,
  \dodoi{10.1016/j.icarus.2010.07.013}

\bibitem[{{Zhang}(2020)}]{Zhang2020}
{Zhang}, X. 2020, Research in Astronomy and Astrophysics, 20, 099,
  \dodoi{10.1088/1674-4527/20/7/99}

\bibitem[{{Zhang} \& {Showman}(2014)}]{zhang14}
{Zhang}, X., \& {Showman}, A.~P. 2014, \apjl, 788, L6,
  \dodoi{10.1088/2041-8205/788/1/L6}

\bibitem[{{Zhou} {et~al.}(2020{\natexlab{a}}){Zhou}, {Bowler}, {Morley},
  {Apai}, {Kataria}, {Bryan}, \& {Benneke}}]{Zhou2020b}
{Zhou}, Y., {Bowler}, B.~P., {Morley}, C.~V., {et~al.} 2020{\natexlab{a}}, \aj,
  160, 77, \dodoi{10.3847/1538-3881/ab9e04}

\bibitem[{{Zhou} {et~al.}(2018){Zhou}, {Apai}, {Metchev}, {Lew}, {Schneider},
  {Marley}, {Karalidi}, {Manjavacas}, {Bedin}, {Cowan}, {Miles-P{\'a}ez},
  {Lowrance}, {Radigan}, \& {Burgasser}}]{Zhou2018}
{Zhou}, Y., {Apai}, D., {Metchev}, S., {et~al.} 2018, \aj, 155, 132,
  \dodoi{10.3847/1538-3881/aaabbd}

\bibitem[{{Zhou} {et~al.}(2019){Zhou}, {Apai}, {Lew}, {Schneider},
  {Manjavacas}, {Bedin}, {Cowan}, {Marley}, {Radigan}, {Karalidi}, {Lowrance},
  {Miles-P{\'a}ez}, {Metchev}, \& {Burgasser}}]{Zhou2019}
{Zhou}, Y., {Apai}, D., {Lew}, B. W.~P., {et~al.} 2019, \aj, 157, 128,
  \dodoi{10.3847/1538-3881/ab037f}

\bibitem[{{Zhou} {et~al.}(2020{\natexlab{b}}){Zhou}, {Apai}, {Bedin}, {Lew},
  {Schneider}, {Burgasser}, {Manjavacas}, {Karalidi}, {Metchev},
  {Miles-P{\'a}ez}, {Cowan}, {Lowrance}, \& {Radigan}}]{Zhou2020}
{Zhou}, Y., {Apai}, D., {Bedin}, L.~R., {et~al.} 2020{\natexlab{b}}, \aj, 159,
  140, \dodoi{10.3847/1538-3881/ab6f65}

\bibitem[{{Zink} {et~al.}(2020){Zink}, {Hardegree-Ullman}, {Christiansen},
  {Petigura}, {Dressing}, {Schlieder}, {Ciardi}, \& {Crossfield}}]{Zink2020}
{Zink}, J.~K., {Hardegree-Ullman}, K.~K., {Christiansen}, J.~L., {et~al.} 2020,
  \aj, 160, 94, \dodoi{10.3847/1538-3881/aba123}

\end{thebibliography}
\bibliographystyle{aasjournal}

\end{CJK*}
\end{document}